\documentclass{jfm}
\usepackage{graphicx}
\usepackage{epstopdf, epsfig}
\usepackage{xcolor}
\usepackage{soul}
\usepackage{amsmath}

\shorttitle{Dynamics and evolution of Turbulent Taylor rolls}
\shortauthor{F. Sacco, R. Verzicco and R. Ostilla-M\'onico}

\title{Dynamics and evolution of Turbulent Taylor rolls}

\author{Francesco Sacco\aff{1}
  \corresp{\email{francesco.sacco@gssi.it}},
  Roberto Verzicco\aff{1,2,3}
 \and Rodolfo Ostilla-M\'onico\aff{4}\corresp{\email{rostilla@central.uh.uk}} }

\affiliation{\aff{1}Gran Sasso Science Institute, Viale Francesco Crispi 7, L’Aquila 67100, Italy
\aff{2} Dipartimento di Ingegneria Industriale, University of Rome ``Tor Vergata", Via del Politecnico 1, Roma 00133, Italy
\aff{3} Physics of Fluids Group, Faculty of Science and Technology, MESA+ Research             Institute, and J. M. Burgers Centre for Fluid Dynamics, University of Twente, PO Box 217, 7500 AE Enschede, The Netherlands
\aff{4} Cullen College of Engineering, University of Houston, Houston, TX 77204, USA

}

\begin{document}

\maketitle

\begin{abstract}
In many shear- and pressure-driven wall-bounded turbulent flows secondary motions spontaneously develop and their interaction with the main flow alters the overall large-scale features and transfer properties. Taylor-Couette flow, the fluid motion developing in the gap between two concentric cylinders rotating at different angular velocity, is not an exception, and toroidal Taylor rolls have been observed from the early development of the flow up to the fully turbulent regime. In this manuscript we show that under the generic name of ``Taylor rolls'' there is a wide variety of structures that differ for the vorticity distribution within the cores, the way they are driven and their effects on the mean flow. We relate the rolls at high Reynolds numbers not to centrifugal instabilities, but to a combination of shear and anti-cyclonic rotation, showing that they are preserved in the limit of vanishing curvature and can be better understood as a pinned cycle which shows similar characteristics as the self-sustained process of shear flows. By analyzing the effect of the computational domain size, we show that this pinning is not a product of numerics, and that the position of the rolls is governed by a random process with the space and time variations depending on domain size. 
\end{abstract}


\section{Introduction}

Taylor-Couette (TC) flow, the fluid flow between two coaxial, independently rotating cylinders, is one of the paradigmatic systems of fluid dynamics, and a model system for studying the influence of rotation and curvature on turbulence \citep{gro16}. Historically, this system has been heavily researched from the 1890s up to the present day, uncovering interesting phenomena for a system that even at the lowest Reynolds numbers is extremely rich. Taylor-Couette flow is linearly unstable when angular momentum decreases with radius, and different configurations of inner and outer cylinder rotation lead to diverse dynamics \citep{and86}. Large-scale structures arise because of instabilities and can fill the entire gap between the cylinders remaining relatively stationary in time. These structures, known as Taylor rolls or Taylor vortices \citep{tay23}, have been observed both in experiments and simulations, and dominate the flow topology up to the highest values of Reynolds number \citep{lat92,hui14,ost14}; here the Reynolds number $Re$ is defined as $Re=Ud/\nu$, with $U$ a characteristic shear velocity of the cylinders which will be defined, $d$ the distance between the cylinders and $\nu$ the kinematic viscosity.  

The phase diagram of Taylor rolls is quite complex, and strongly dependent on the kind of rotation of the two cylinders. In the case of outer cylinder rotation and for a fixed or only slowly rotating inner cylinder, Taylor-Couette flow is linearly stable. On the other hand for pure inner cylinder rotation the centrifugal driving forces make the system unstable as the Reynolds number of the inner cylinder increases. At Reynolds numbers just beyond the onset of linear instability ($Re\sim 10^2$),  Taylor rolls appear and are stationary and axisymmetric \citep{and86}.  For increasing Reynolds number, secondary instabilities arise, and the Taylor rolls develop azimuthal oscillations, entering the wavy Taylor Vortex flow regime \citep{jon85}. Further increasing the Reynolds number causes the onset of time-dependence, and this is known as the modulated wavy Taylor Vortex regime. When increasing the driving even more, even if the large coherent rolls are still present, the flow becomes chaotic as turbulence starts to develop, first in the bulk, and eventually also in the boundary layers as they undergo a shear transition \citep{ost14}. This is the so-called turbulent Taylor vortex regime. For Reynolds numbers exceeding $Re\sim10^3$, the small-scale vortices begin to be strong enough to weaken the large-time-scale circulation. For pure inner cylinder rotation, \cite{lat92} found experimentally that the rolls disappear for $\eta=0.714$ and $Re>10^5$, where $\eta=r_i/r_o$ is the radius ratio of the two cylinders, with $r_i$ ($r_o$) the inner (outer) cylinder radius. When the Reynolds number becomes sufficiently large, the flow recovers a statistical symmetry in the axial direction. The vanishing of the rolls at $\eta=0.714$ was also found numerically in \cite{ost14}, albeit for smaller aspect ratios $\Gamma=L_z/d$, where $L_z$ is the axial length of the domain, $d$ is the gap between the two cylinders.

When both cylinders rotate, the dynamic of the rolls becomes more complicated. By adding a slight counter-rotation to a $\eta=0.714$ TC system, \cite{hui14} found that the turbulent Taylor rolls survive at least up to Re $\sim 10^6$. Numerically, \cite{ost14b} observed rolls in the corresponding counter-rotation parameter regimes in DNS simulations for $\eta=0.714$ and $\Gamma=2.07$, from the formation of the axisymmetric rolls at low Reynolds number, up to the turbulent Taylor rolls. \cite{ost14b} also explored the phase space for the rolls in the high Reynolds number regime, showing that they can either fill the entire gap, partially survive close to the inner cylinder or completely disappear from the bulk  depending on rotation and curvature of the system. Even if a smaller aspect ratio was used, wide agreement was found between the experimental and numerical probes of the counter-rotating regime.

The transition between two stages of the phase diagram in our previous analysis is far from smooth, and the continuity we have assumed between the steady, axisymmetric vortices near the onset of centrifugal instability to the large-$Re$ turbulent Taylor rolls is not totally clear. Several questions remain: first, Taylor rolls have been traditionally attributed to the effects of centrifugal instabilities. However, for pure inner cylinder rotation, they appear to be strongest for narrow-gap (i.e.~small curvature) instances of Taylor-Couette flow \citep{ost14}. For high curvature ($\eta=0.5$), they survive only at mild counter-rotation \citep{van16}. If they were caused solely by centrifugal effects, one could expect them to be always the strongest for the largest centrifugal driving, i.e.~pure inner cylinder rotation, and this is not the case. 

In this manuscript we first intend to clarify the reason for this discrepancy by simulating small-gap Taylor-Couette flow at moderate Reynolds numbers and varying the curvature. By approaching TC flow with zero curvature, and analyzing its transition to rotating plane Couette flow (RPCF), the flow between two infinite and parallel plates moving at different velocities, we will study how the linear instabilities of TC flow relate to those of processes only driven by shear. This idea has already been used in the literature to study solutions of plane Couette flow: by the use of homotopy, \cite{nag90} was able to continuously deform a low Reynolds number Taylor vortex from a TC solution for counter-rotating cylinders into a narrow gap onto a RPCF solution; \cite{fai00} studied the narrow gap limit of TC flow, showing how the linear instability that leads to the formation of Taylor vortices is superseded by the planar shear flow transition as the curvature vanishes.

In a shear-driven flow, e.g.~plane Couette flow, with linear instabilities absent, turbulence is regenerated by the self-sustaining process (SSP) first characterized by \cite{Wal97}. The cycle, detailed in $\S3.2$, consists of three steps, and each of these is necessary to maintain turbulence \citep{jim91,ham95}. Recently, a relationship between Taylor vortices at low curvature and the SSPs of shear flows was found by \cite{des18}, who showed that at low Reynolds numbers the transition from Taylor vortex flow to wavy-vortex flow is caused by the activation of one of the phases of the SSP: the formation of the streak instability. And once the azimuthal waves arise on the rolls, their non-linear interaction feeds the rolls, closing the SSP. In this manuscript, we intend to show how the flow behaves as the Reynolds number increases and to check if imprints of this behaviour can still be seen. 

A second question that calls for explanation is about the axial pinning of the rolls. While this comes about naturally in experiments due to the presence of end-plates, in direct numerical simulations it is still not clear why structures should be fixed in a homogeneous direction, and the broken symmetry is not recovered in a statistical sense. Several numerical studies have been conducted to check that the axial pinning of the rolls is not an effect of small computational boxes due to aliasing of long-wavelength modes onto the streamwise-invariant mode \citep{ost15,ost16}, finding no apparent effect of box-size on the rolls up to box sizes of $20\pi \delta \times 2\delta \times 9\pi\delta$, with $\delta$ the half-gap. However on one side it has been found that for what would be considered large boxes, i.e.~ the DNS simulations in a domain of size $20\pi\delta \times 2\delta \times6\pi\delta$, with $\delta$ the half-plate distance by \cite{avs14}, or the $18\pi\delta\times2\delta\times8\pi$ domain of \cite{pir14}, no pinned structures appeared. On the other side a recent numerical work by \cite{lee18} has shown that in plane Couette flow structures become pinned in a comparable large computational box of $20\pi\delta \times 2\delta \times5\pi\delta$ for high Reynolds number. 

It thus seems that it is impossible to neglect the effect of computational domain size, and in this paper we set out to show in a more rigorous manner than simply showing visualizations as done previously \citep{ost15,ost16}, how effects of finite domain sizes on structures can be quantified by increasing the azimuthal extent of the simulation domain. 

The paper is organized as follows: in section \ref{sec:numerics} we describe the numerical method used and the parameter setup used in the simulations. All the properties and the characterization of Taylor rolls are reported in section \ref{sec:results}. In section \ref{sec:conclusion} we briefly summarize and give some remarks and an outlook for future works.

\section{Numerical details}\label{sec:numerics}

We perform direct numerical simulations of Taylor-Couette flow by solving the incompressible Navier-Stokes equations in a rotating reference frame:

\begin{equation}
 \displaystyle\frac{\partial \textbf{u}}{\partial t} + \textbf{u}\cdot \nabla \textbf{u} + 2\Omega\times\textbf{u} = -\nabla p + \nu \nabla^2 \textbf{u},
 \end{equation}
 
 \begin{equation}
  \nabla \cdot \textbf{u} = 0,
 \end{equation}

\noindent in cylindrical coordinates, where $\textbf{u}$ is the velocity, $\Omega$ the angular velocity of the rotating frame, $p$ the pressure and $t$ is time. Spatial discretization is performed using a second-order energy-conserving centered finite difference scheme, while time is advanced using a low-storage third-order Runge-Kutta for the explicit terms and a second-order Adams-Bashworth scheme for the implicit treatment of the wall-normal viscous terms. Further details of the algorithm can be found in \cite{ver96,poe15}. The code has been extensively validated for Taylor-Couette flow \citep{ost14} and is parallelized using MPI directives. 

Axially periodic boundary conditions are taken with a periodicity length $L_z$, expressed non-dimensionally as an aspect ratio $\Gamma=L_z/d$, where $d=r_o-r_i$ is the gap between both cylinders, and $r_i$ ($r_o$) the radius of the inner (outer) cylinder. For this study, we fix $\Gamma=2.33$, which is enough to fit a single roll pair, so that their wavelength $\lambda_z$ will be $\lambda_z=\Gamma=2.33$. The non-dimensional radius ratio $\eta=r_i/r_o$ provides a measure of the curvature of the system, and is the second geometrical control parameter. In the azimuthal direction, Taylor-Couette is naturally periodic. However, a rotational symmetry of order $N$ is imposed on the system with a double purpose: to reduce computational costs and to explore the effect of the the streamwise periodic length on turbulent structures. In this study, we mainly consider rotational symmetries between $5$ and $20$. For $\eta=0.909$, a rotational symmetry of $N=20$ results in a streamwise periodicity length of around $2\pi$ half-gaps, the usual value considered for the turbulent channel flow. We note that while \cite{and86} found azimuthal waves of wavenumber $m=2$ near the transition to wavy Taylor vortex flow at similar radius ratio, the moderate Reynolds number simulations of \cite{ost15} show very rapid drops in the streamwise decorrelation length. Indeed, $N=20$ was found to be enough for pure inner cylinder rotating Taylor-Couette in \cite{ost15} and \cite{ost16} to obtain asymptotic torque and mean statistics, compared to the largest extents of $N=5$. 

We perform the simulations in a convective reference frame \citep{dub05} such that the cylinders rotate with opposite velocities $\pm U/2$, and any combination of differential rotations of the cylinders are  reflected as a Coriolis force. In this frame the two control parameters become a shear Reynolds number $Re_s=Ud/\nu$ and a non-dimensional Coriolis parameter $R_\Omega=2d\Omega /U$; if we consider the
traditional Reynolds numbers that measure the dimensionless velocity of the inner and
outer cylinders in the laboratory frame of reference $Re_i=r_i \omega_i d /\nu$ and $Re_o=r_o \omega_o d /\nu$, where $\omega_i$ and $\omega_o$ are the angular velocities of the inner and outer cylinders, we can express the new control parameters as follows:
\begin{align}
Re_s &= \frac{2}{1 + \eta}(Re_i + \eta Re_o),\\
R_\Omega &= (1-\eta)\frac{Re_i + Re_o}{Re_i - \eta Re_o}.
\end{align}

In order to compare Taylor-Couette system in the limit of a vanishing curvature, also a few simulations of Plane-Couette flow have been performed, since it is the limit of Taylor-Couette flow when $\eta\to 1$. For this case we have used a Cartesian version of the previously mentioned code \citep{poe15}. In the convective frame, the control parameters $Re_s$ and $R_\Omega$ naturally reduce to those of rotating Plane-Couette flow when that limit is taken.

Temporal convergence is assessed by measuring the difference in torque between both cylinders. The flow is started from a white noise configuration, and run for approximately $tU/d=1000$ time units to overcome the transient. The simulations are then advanced until the time average of the respective values are equal within 1\%, which roughly corresponds to an additional $tU/d=200$ time units. The torque is then taken as the average value of the inner and outer cylinder torques. Therefore, the error due to finite time statistics can be estimated to be around 1\%. A complete discussion of temporal time-scales is available in \cite{ost16b}. The spatial resolution of lower $Re_s$ simulations is based on \cite{ost16}. For $Re_s=3.61\cdot 10^4$, we take the criteria $\Delta z^+\approx 5$, $\Delta x^+ = r\Delta \theta^+\approx 9$ and $\Delta r^+\in (0.5,5)$, resulting in resolutions of $n_\theta\times n_r\times n_z = 384 \times 512 \times 768$ in the azimuthal, radial and axial directions respectively for $N=20$. For simulations in which we halve $N$, we double the resolution in the azimuthal direction.

\section{Results}
\label{sec:results}

\subsection{Changes in the rolls with Reynolds number}
\label{subsec:LSE}

\begin{figure}
\includegraphics[trim=4cm 0 5cm 0, clip, height=0.29\textwidth]{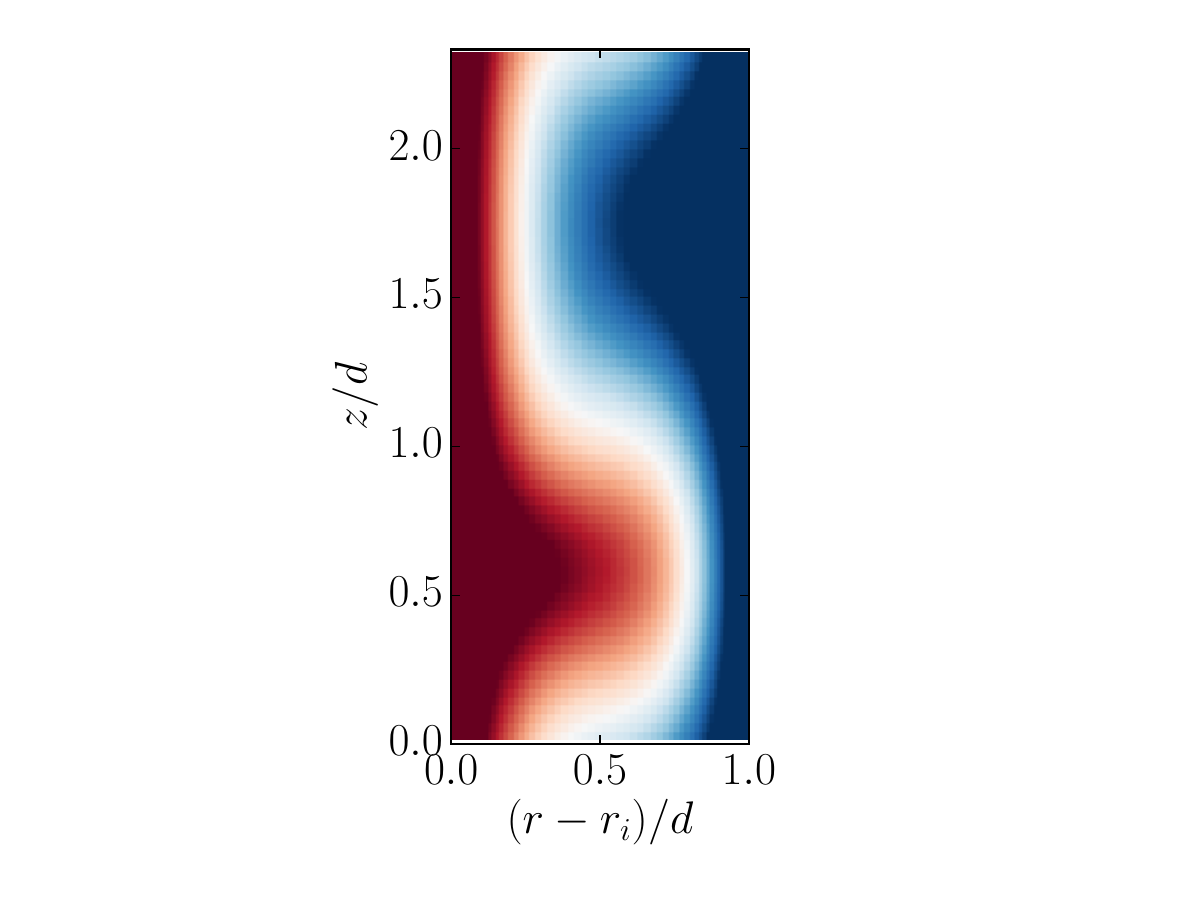}
\includegraphics[trim=4cm 0 5cm 0, clip, height=0.29\textwidth]{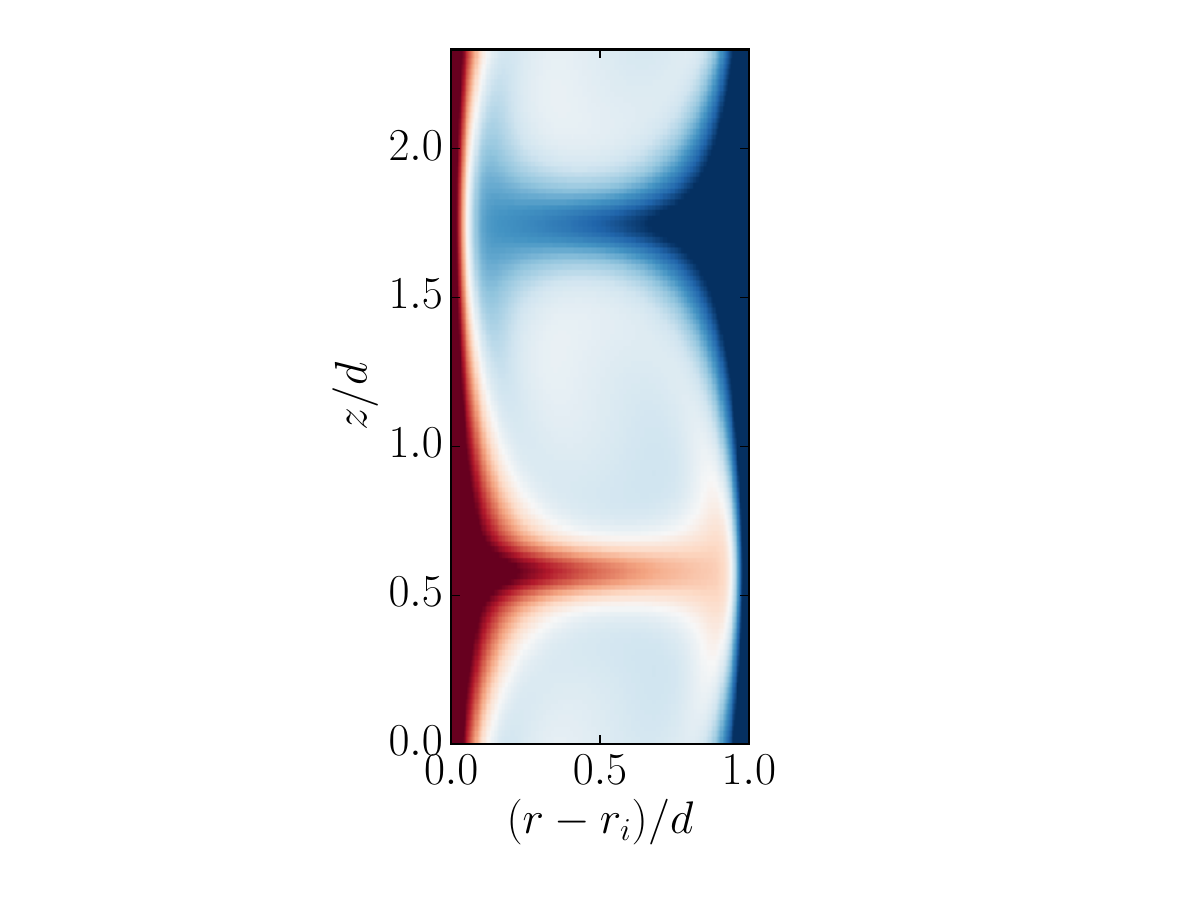}
\includegraphics[trim=4cm 0 5cm 0, clip, height=0.29\textwidth]{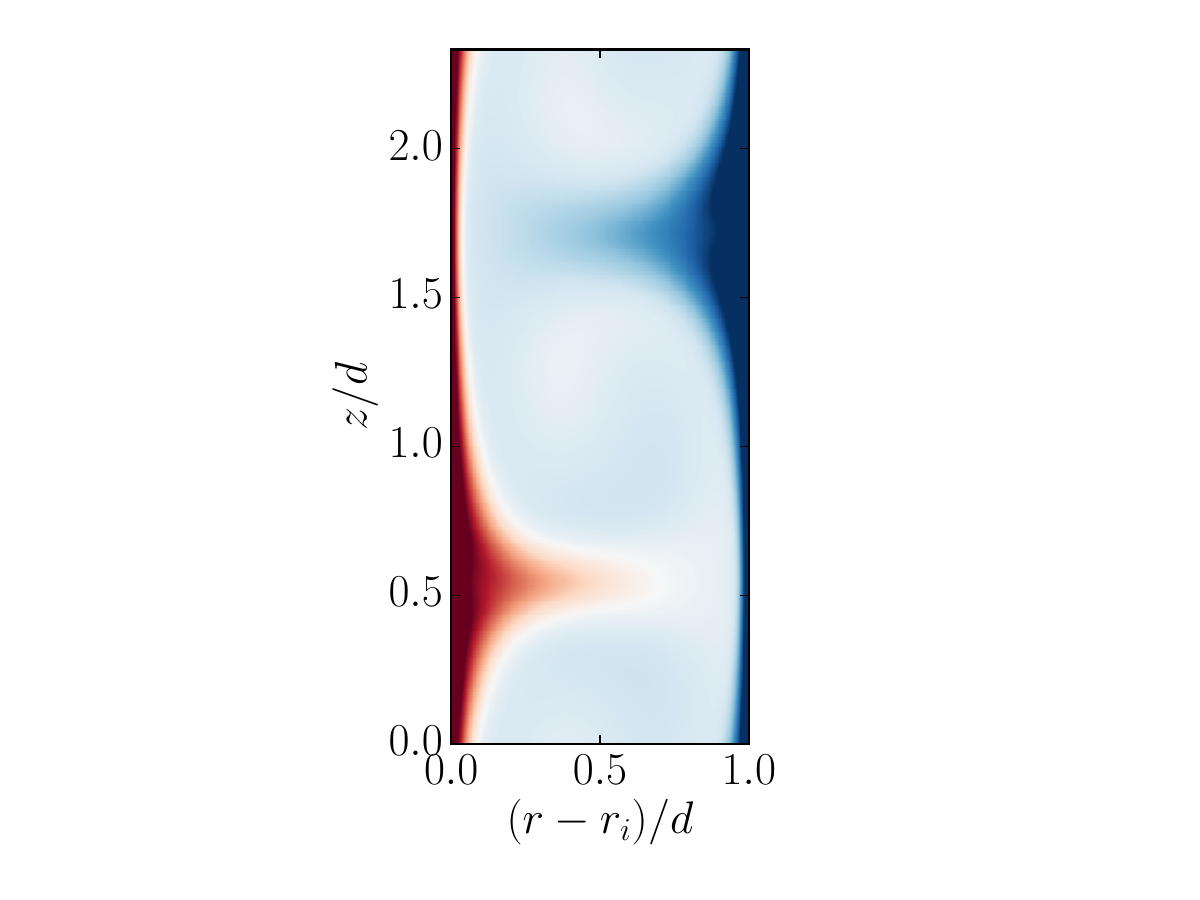}
\includegraphics[trim=6cm 0 0cm 0, clip, height=0.297\textwidth]{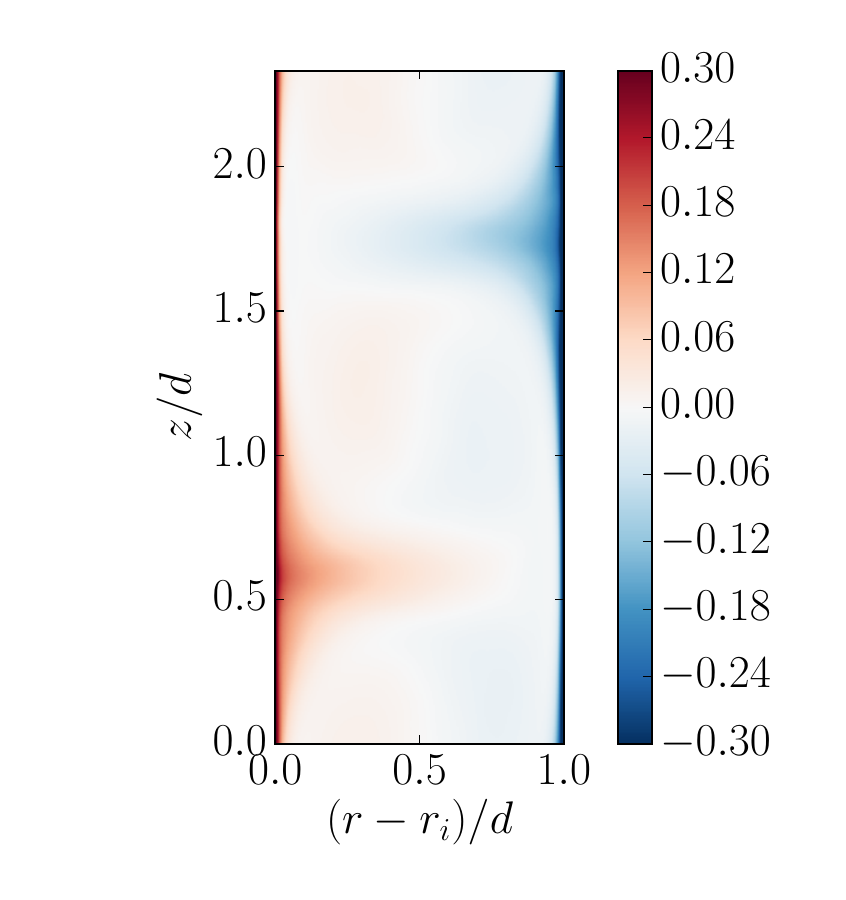}\\
\includegraphics[trim=4cm 0 5cm 0, clip, height=0.3\textwidth]{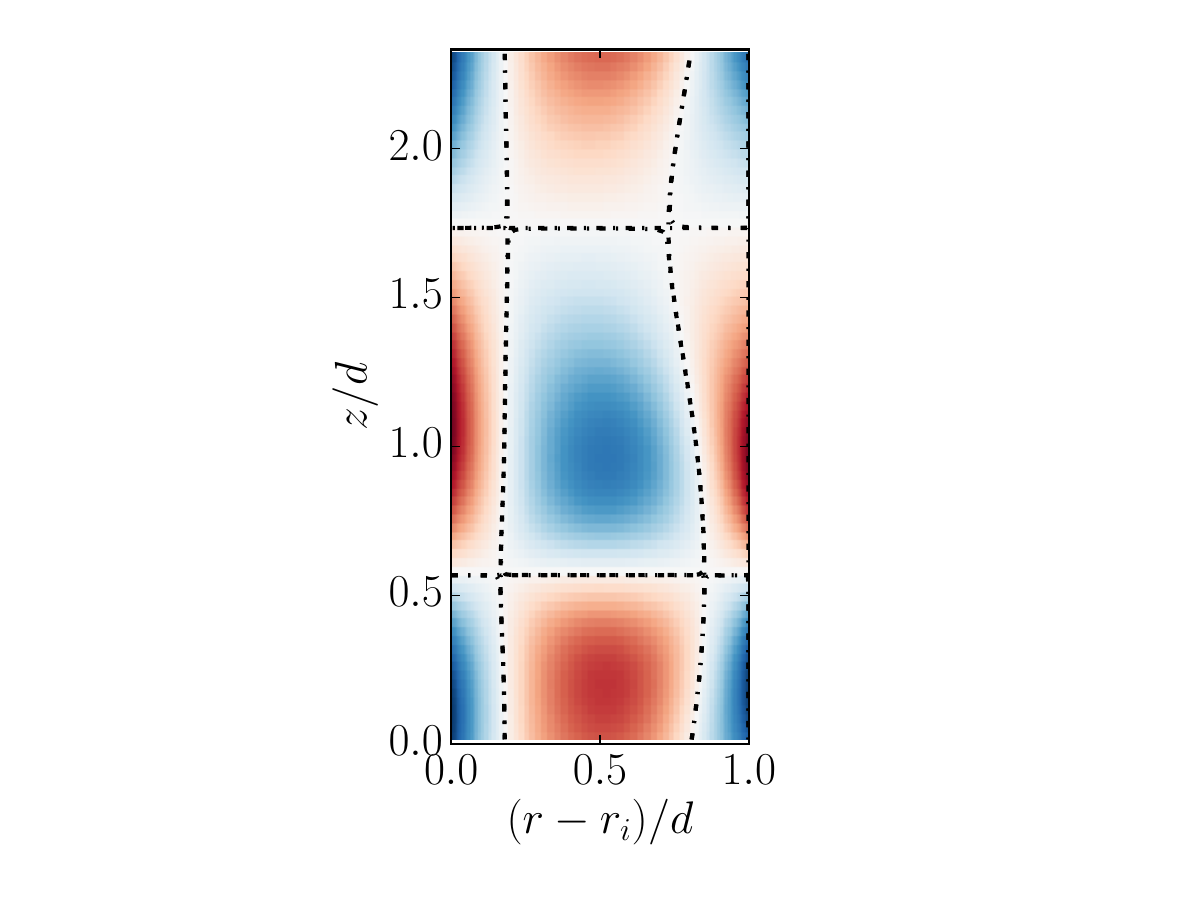}
\includegraphics[trim=4.4cm 0 5cm 0, clip, height=0.3\textwidth]{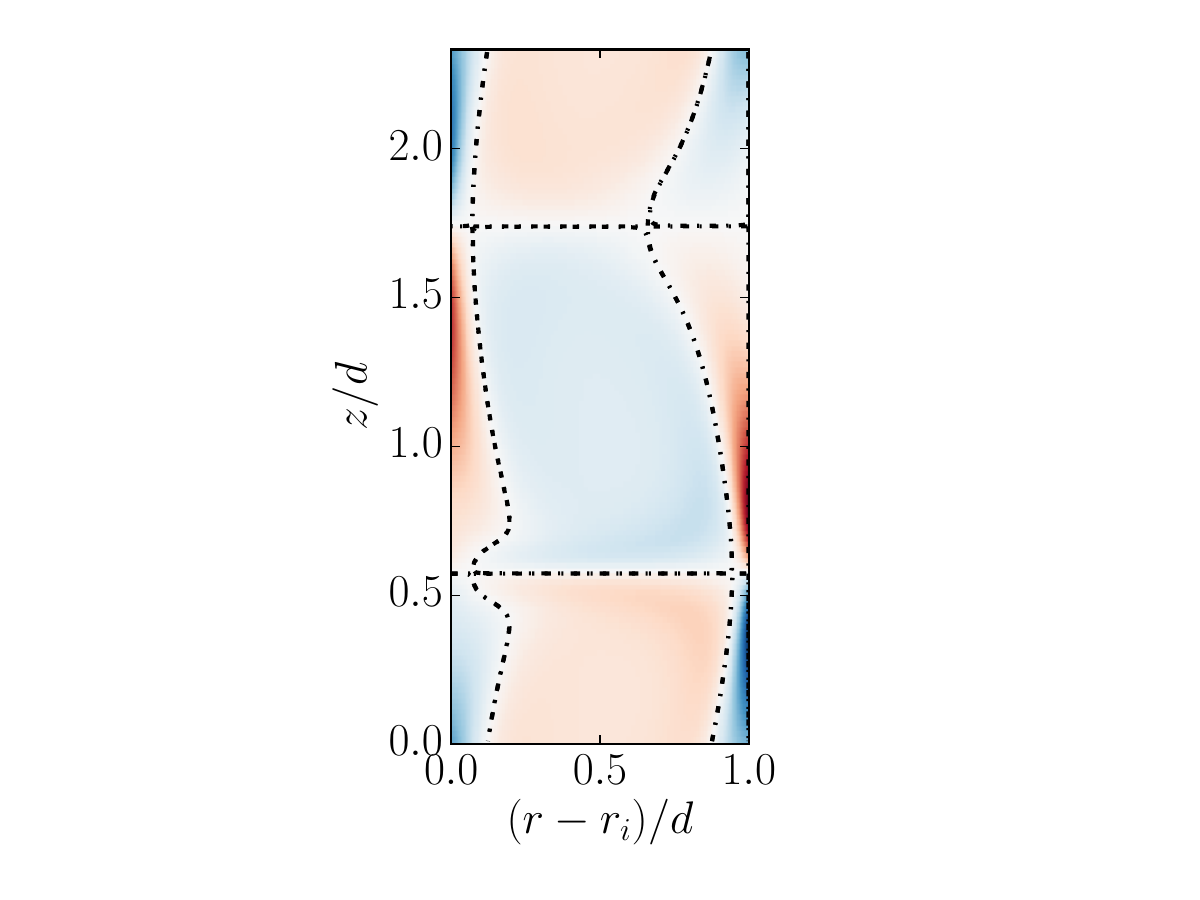}
\includegraphics[trim=4.5cm 0 5cm 0, clip, height=0.3\textwidth]{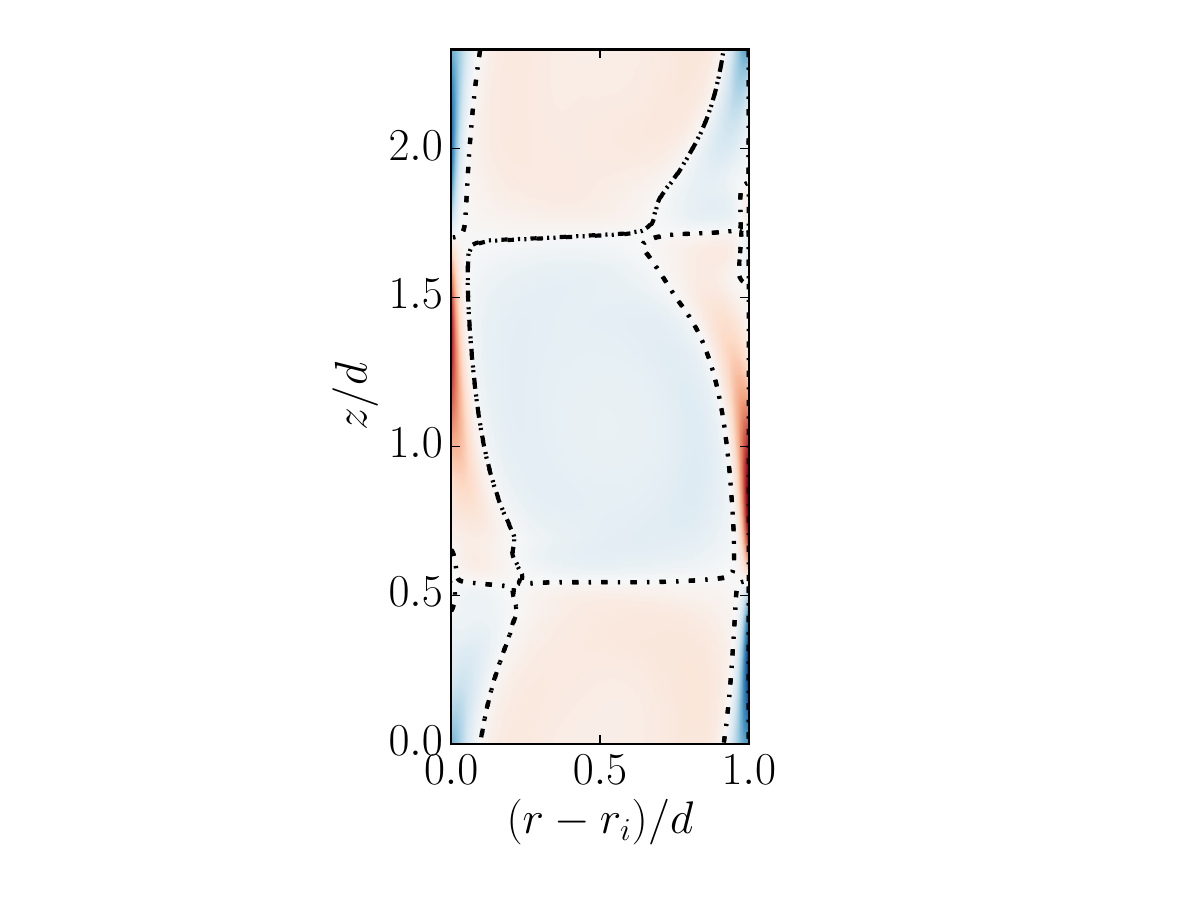}
\includegraphics[trim=6.5cm 0cm 0cm 0cm, clip, height=0.307\textwidth]{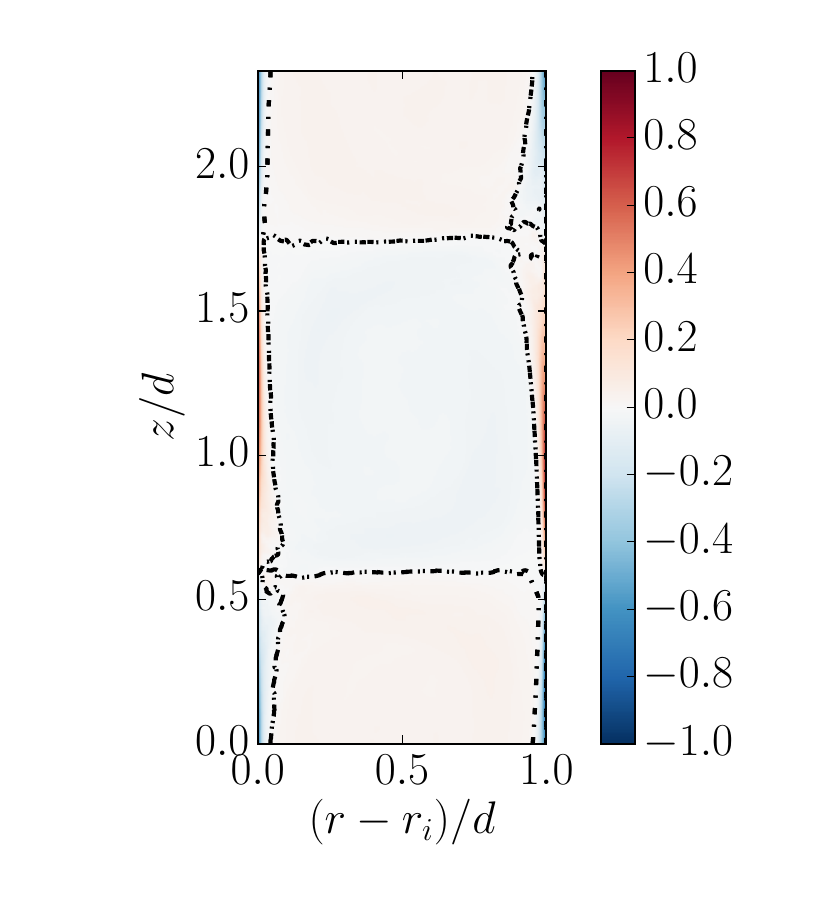}\\
\includegraphics[trim=-1cm 0.2cm 0cm 0cm, clip, height=0.35\textwidth]{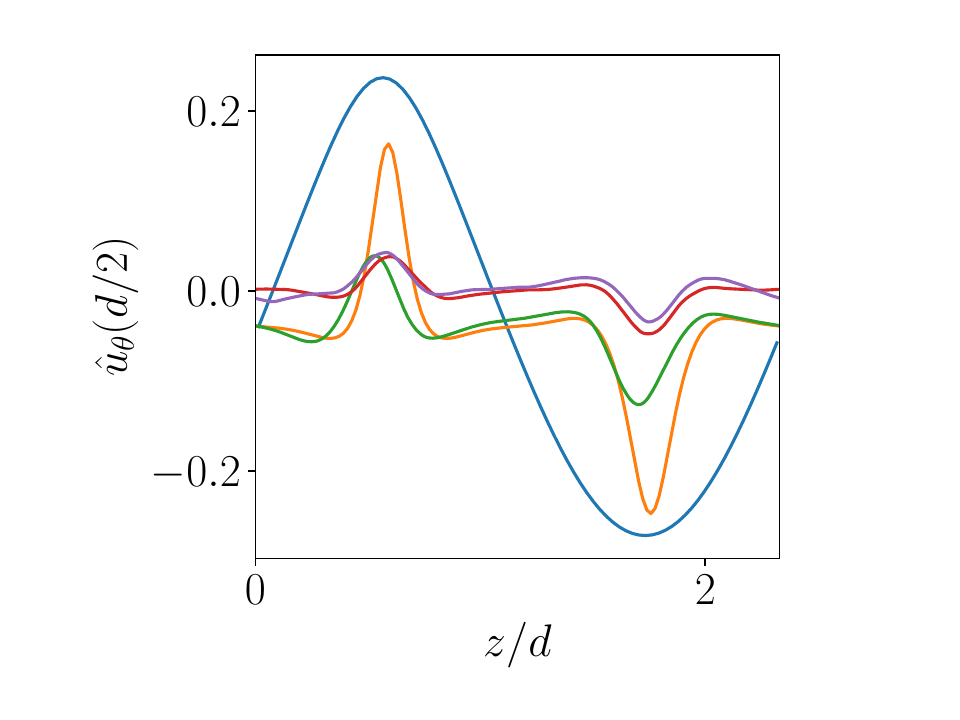}
\includegraphics[trim=1cm 0cm 0cm 0cm, clip, height=0.35\textwidth]{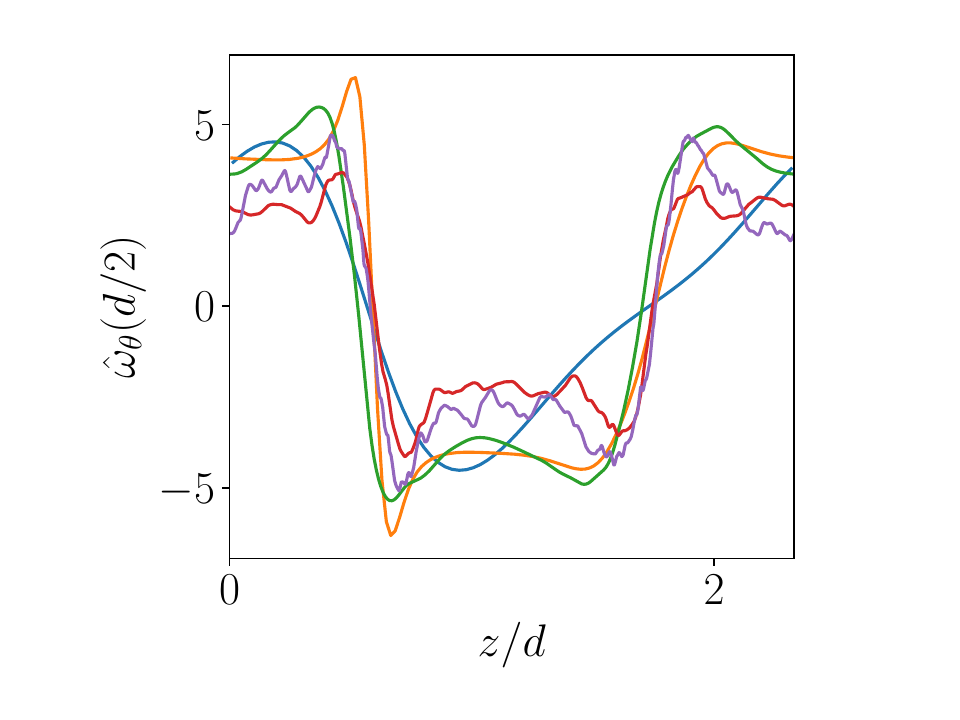}
\\

\centering
\caption{Temporally- and azimuthally averaged azimuthal velocity $\hat{u}_\theta$ (top row) and normalized averaged azimuthal vorticity $\hat{\omega}_\theta/\max |\hat{\omega}_\theta|$ (middle row) for inner cylinder rotation and several shear Reynolds numbers (from left to right, $Re_s=1.96\cdot 10^2$, $1.37\cdot10^3$, $4.97\cdot10^3$, $3.61\cdot10^4$) at $\eta=0.909$ and $\Gamma=2.33$. Contours levels for vorticity are shown for values between -0.001 and 0.001, to highlight the roll structures. The panels on the bottom row show two quantities, azimuthal velocity (left) and azimuthal vorticity (right) at the mid-gap $r=d/2$ for Reynolds numbers  $Re_s=1.96\cdot 10^2$ (blue), $1.37\cdot10^3$ (orange), $4.97\cdot10^3$ (green), $3.61\cdot10^4$ (red) and $1 \cdot 10^5$ (purple).}  
\label{fi:avgvelsvorts} 
\end{figure}

In order to analyze the properties of the large scale structures, we have performed several simulations fixing the cylinder radius ratio to $\eta=0.909$, and increasing the driving by rotating the inner cylinder only and varying the Reynolds number $Re_s$ between $10^2$ and $3.6\cdot10^4$, thus covering the full range between linear instability and turbulent rolls. Indeed, for all the cases, the axially pinned structures exist and fill up the entire domain. These Taylor rolls are known to redistribute angular momentum and have a visible large scale pattern \citep{lat92,ost16b}, as shown in the first two rows of Figure \ref{fi:avgvelsvorts}: if we look at the temporally and azimuthally averaged azimuthal velocity and azimuthal vorticity, indeed, we can see that the signature of the rolls is preserved for all cases. Following \cite{bus12}, Taylor rolls were analyzed with the terminology of thermal convection by \cite{ost14b}, where it was shown that at both cylinder walls, there existed three types of areas: plume ejection regions, where the velocity was largely in the positive wall-normal direction, and turbulent plume-like structures would leave the cylinder, plume impact regions, where the velocity was largely in the negative wall-normal direction, and plume-like structures from the other cylinder would impact the wall, and wind-shear regions where the velocity was largely parallel to the wall, and plumes would be generated. In \cite{ost13,ost14b}, the plumes were related to herringbone-structures, or hairpin vortices, which can also be seen in the left panel of Figure \ref{fi:inststreaks}.

In Figure \ref{fi:avgvelsvorts}, the two quantities show different behaviours: the mean azimuthal velocity preserves a similar pattern during all the transition, revealing the Taylor rolls and the two zones where the fluid detaches from the inner/outer boundary layer and impacts the opposite wall. Alternatively, the intensity of the azimuthal (streamwise) vorticity concentrates near the wall. While we could expect the axial (spanwise) vorticity to concentrate near the walls with increasing Reynolds number due to boundary layer thinning, the behaviour of the azimuthal vorticity is more complicated. If there were no secondary flow, one could expect the mean azimuthal vorticity to average out to zero. In contrast, if secondary flows were present, as for example in the case of square-duct flow \citep{pir18}, azimuthal vorticity would be largely concentrated not only in the boundary layers but also in the \emph{core} of the secondary structures. For Taylor-Couette flow with inner cylinder rotation, the azimuthal vorticity in the boundary layer becomes much higher than that of the core of the secondary flow, i.e. the Taylor roll. For this to happen, the mean flow velocity of the secondary motion must be relatively constant with Reynolds number, and the boundary layer of the secondary motion must become thinner. This results in a relative ``emptying'' process of the bulk of the rolls. It is possible to see this process in the last row of Figure \ref{fi:avgvelsvorts}, where in the right panel we have checked the magnitude of the mean streamwise vorticity in the core of the rolls, i.e. at the mid-gap, along the spanwise coordinate. We can clearly see that, while at low Reynolds number of $Re=1.96 \cdot 10^2$ the peak is located in the very core of the roll, already during the transition to ultimate regime the center of the structure is ``deflating'', and vorticity starts to concentrate at the borders of the structure. This condition is enhanced at the highest Reynolds numbers $Re=3.61 \cdot 10^4$ and $Re=1 \cdot 10^5$, where the relatively constant value of azimuthal vorticity drops. This fact shows that other processes are taking place. As a consequence rolls become relatively more quiescent and inactive even if some circulation remains in the secondary flow cores.

This matches previous intuition that as the Reynolds number increases and the flow reaches the ultimate regime, the rolls undergo a series of transitions. In \cite{ost14b}, when the ultimate-regime was reached for pure inner cylinder rotation at $\eta=0.714$, the rolls simply vanished. For $\eta=0.909$, even if they persist, between the right-most upper panels at $Re_s=3.61\cdot 10^4$, corresponding to the so-called ultimate regime, and the panels at $Re_s=4.97\cdot 10^3$, the azimuthal velocity in the bulk can be seen to increase even if strong axial signatures are still present. The left panel of the last row of Figure \ref{fi:avgvelsvorts}, representing the magnitude of the mean streamwise velocity in the core of the rolls, i.e. at the mid-gap, along the spanwise coordinate, is a confirmation of this fact: even if, as we have already said, the profile of each velocity preserves the concentrated shape due to fluid detaching/impacting pattern, there is a velocity value jump in correspondence with the transition to the ultimate regime, between $Re_s=4.97\cdot 10^3$ and $Re_s=3.61\cdot 10^4$, that somehow matches the drop of azimuthal vorticity in the core seen in the right panel. 

This calls into question the assumption that turbulent Taylor rolls before the ultimate regime such as those in \cite{and86}, and those in the ultimate regime such as those in \cite{hui14} are generated by the same mechanisms. The continuity between the left and right upper panels becomes less apparent upon close examination, and it appears that fundamental changes happen as the rolls are becoming relatively empty of vorticity.

\subsection{The Taylor roll in the context of the self sustained process}

\begin{figure}
\includegraphics[width=0.48\textwidth]{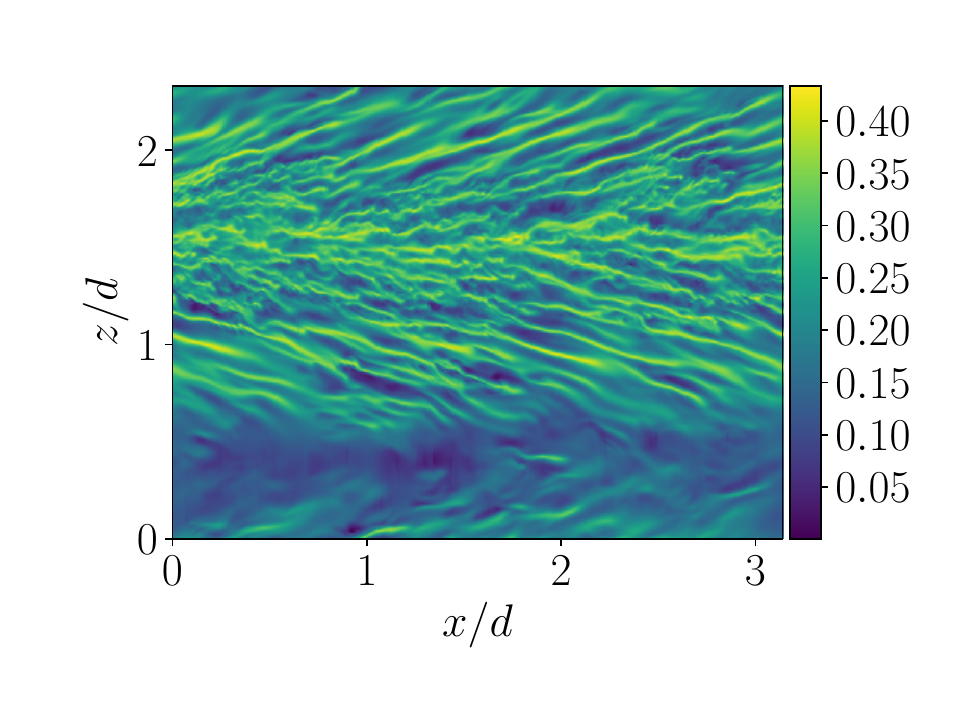}
\includegraphics[width=0.48\textwidth]{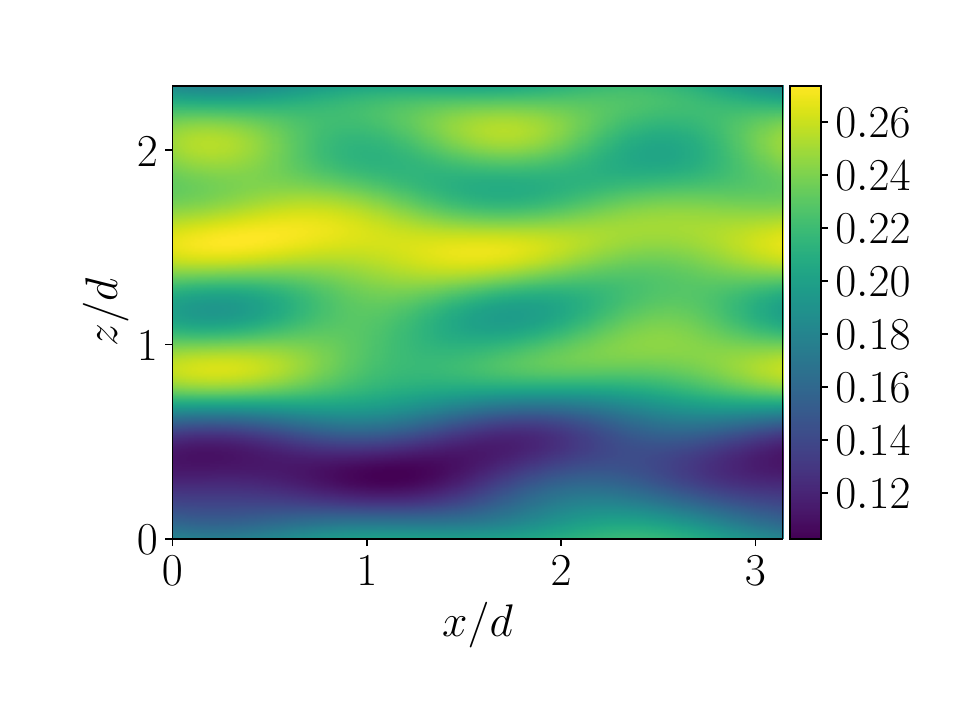}
\centering
\caption{Left: Instantaneous visualization of the azimuthal (streamwise) velocity at $r^+\approx 13$, for $Re_s=3.4 \cdot 10^4$, $\eta=0.909$ and $\Gamma=2.33$. Right: Same velocity field as on the left under a low-pass filter. The characteristic undulatatory streamwise instability of the streaks of the self-sustained process can be clearly seen. }
\label{fi:inststreaks} 
\end{figure}

It thus seems useful to analyze the turbulent Taylor rolls in the ultimate regime not in the context of centrifugal instabilities, but instead in the framework of shear flows. For this aim we introduce the self-sustained process of shear flows. Following the argument of \cite{Wal97}, we know that a self sustained process is composed of three main phases:

\begin{itemize}
\item[i)] the redistribution of mean shear stress by streamwise rolls to create streaks;
\item[ii)] the wake-like instability of the streaks;
\item[iii)] the regeneration of the streamwise rolls from the nonlinear development of the streak instability.
\end{itemize}

In the previous section, we saw that Taylor rolls are essentially large-scale streamwise rolls which redistribute mean shear-stress. \cite{des18} showed that already at low Reynolds numbers, the axisymmetric Taylor roll was part i) of the self-sustained process, and once the transition to wavy Taylor vortices took place, all three parts of the process were active. With increasing Reynolds numbers, we expect the centrifugal (linear) instabilities to become bypassed by the shear instabilities \citep{fai00,bra17}. Thus for high Reynolds numbers we can expect aspects of the Taylor rolls to be related to parts i) and iii) of the SSP. We expect large-scale streaky structures to also exist for Taylor-Couette flow even in the turbulent regime. These structures would then feed the Taylor rolls and axially pinned Taylor rolls would simply be a fixed instance of the above self-sustained process. 

A streaky flow would contain strong spanwise inflections. In the near-wall region of turbulent flows this will lead to two possible patterns, a typical staggered row of vortices and a less frequent horseshoe structure. To detect the existence of these spanwise variations, we have checked the instantaneous azimuthal velocity near the wall for pure inner cylinder rotation, and the results are illustrated in figure \ref{fi:inststreaks}. After applying a low-pass filter to the fields, the streaks-row structure clearly comes up. This provides an initial indication that a Taylor roll in the ultimate regime is an axially pinned process reminiscent of the self-sustained process of shear flows.

\begin{figure}
\includegraphics[width=0.45\textwidth]{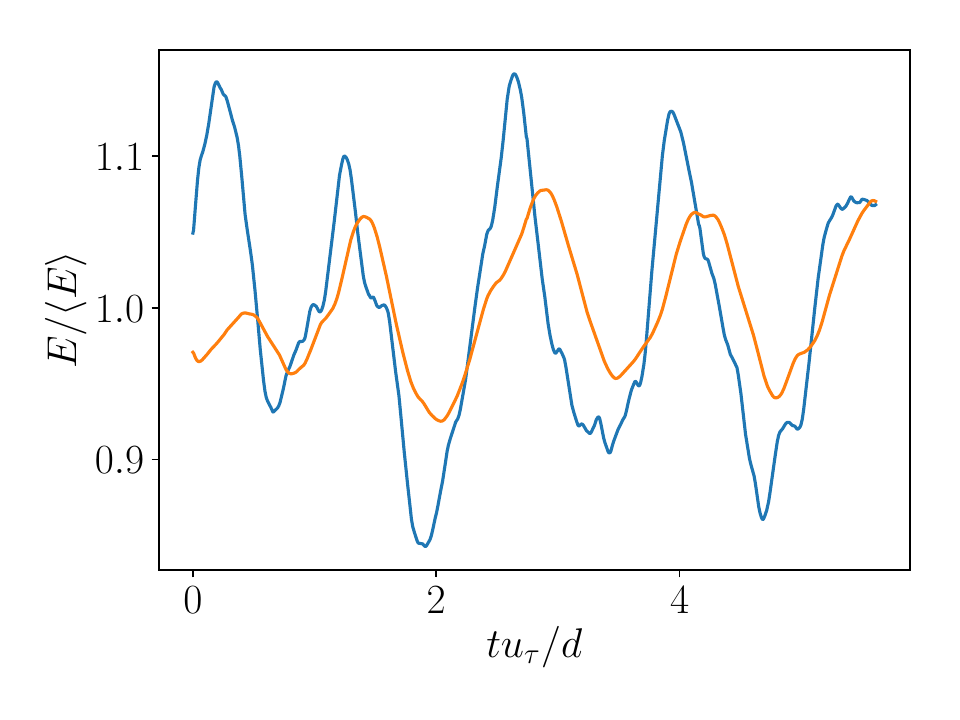}
\includegraphics[width=0.45\textwidth]{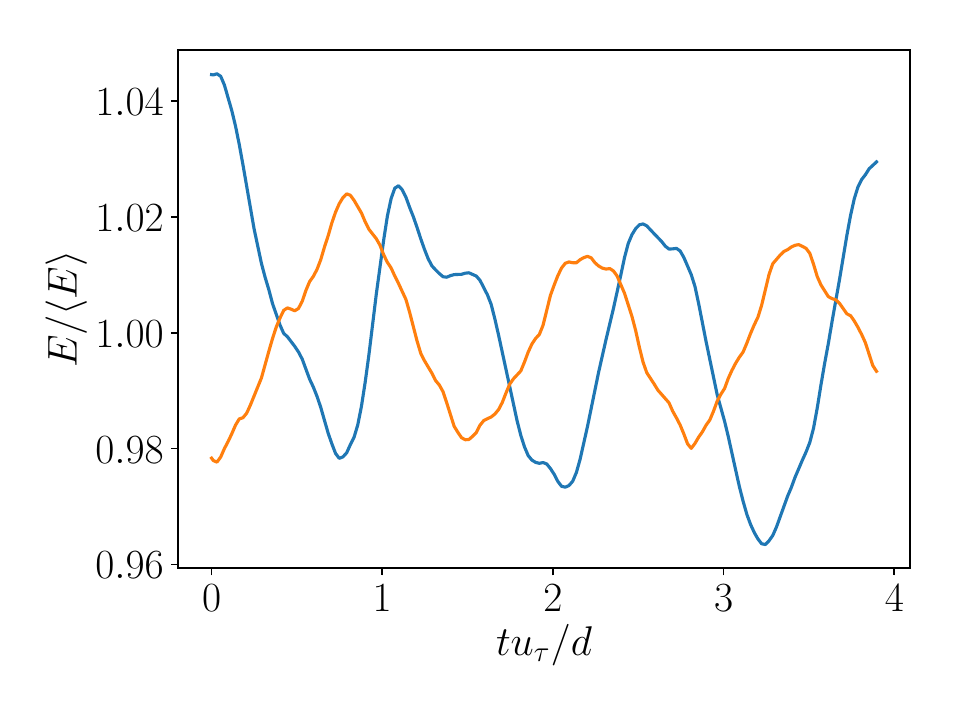}
\centering
\caption{Left: Temporal evolution of the kinetic energy associated to streaks $E_s$ (blue) and rolls $E_r$ (orange) normalized by their mean value. Right: Temporal evolution of the normalized modal RMS velocity in the Fourier space associated to the $M(0,\beta)$ mode (blue) and the $M(0,0)$ mode (orange). Both plots shown are for pure inner cylinder rotation with $\eta=0.909$ and $Re_s=3.4\cdot10^4$.} 
\label{fi:tevo_streak_roll} 
\end{figure}

A further assessment of the similarity between the two processes can be made by studying the temporal evolution of streaks and rolls. There are several definitions used in the literature to quantify the energy of streaks and rolls. First, we following the definition given by \cite{des18} of streaks and rolls in Taylor-Couette flow where their energy is calculated from the energy in axisymmetric modes. 

For this, first of all we separate the field $u$ from its mean $U$, averaged with respect to the streamwise coordinate $\theta$:

\begin{equation}
u(t,\theta,r,z)=U(t,r,z) + u^{'}(t,\theta,r,z),
\label{eq:axisym}
\end{equation}

\noindent and we take a further step by analyzing two derived quantities, the energy in the rolls: 

\begin{equation}
E_r(t,\theta,r,z)=\frac{1}{2} \displaystyle\int_{r,z} [U_r^2(t,r,z) + U_z^2(t,r,z)] r dr dz,
\end{equation}

\noindent and the energy in the streaks as defined by \cite{des18}, which are defined through the axial anomaly, computed taking out the $\theta$ and $z$ average of the field from $U$,

\begin{equation}
E_s(t) = \frac{1}{2} \displaystyle\int_{r,z} [U_\theta(t,r,z) - \left\langle U_\theta(t,r,z) \right\rangle_{z}]^2 r dr dz.
\label{eq:straxi}
\end{equation}

\noindent This is shown in the left panel of Figure \ref{fi:tevo_streak_roll}, where the energy of both structures can be seen to oscillate at time-scales of $\mathcal{O}(u_\tau/d)$. The energy in the rolls can be seen to consistently lag behind the energy in the streaks by a small time, and this is probably due to the fact that they are regenerated through something analogous to step (iii) of the SSP cycle.

A second method to quantify the energy of streaks and rolls, and the link between the rolls and the SSP process is done following \cite{ham95}. Here, the study of streaks and rolls is done through the modal r.m.s.~velocity (or the square root of the `kinetic energy'), in the Fourier space, given by:

\begin{equation}
    M\left(k_{\theta}=\alpha m,k_z=\beta n\right)=\left[ \displaystyle\int_{r_i}^{r_o} \left( \widehat{u_\theta^2}(m \alpha,r,n \beta) +\widehat{u^2_r}(m \alpha,r,n \beta)+ \widehat{u^2_z}(m \alpha,r,n \beta) \right) r dr \right]^{\frac{1}{2}},
\end{equation}

\noindent where $\alpha=2 \pi/L_x$ and $\beta=2\pi/\lambda_z$ are the fundamental streamwise and spanwise wavenubers, $L_x$ corresponding to the streamwise domain extent, and $\lambda_z$ as defined in section \ref{sec:numerics}.

In particular we focus on the two principal modes:
\begin{itemize}
\item $M(0,0)$, the axially and streamwise invariant mode, which represents the mean flow.
\item $M(0,\beta)$ the streamwise independent, fundamental in z mode, corresponding to both streaks and rolls, as both $u_\theta$ and $u_r$/$u_z$ add to the total energy in $M$.
\end{itemize}

In the right panel of Figure \ref{fi:tevo_streak_roll} we see that the two energies oscillate at time-scales of $\mathcal{O}(u_\tau/d)$, and the period of the two quantities are almost anti-correlated, instead of the time-lag of the left panel. This is consistent with a breakdown-regeneration structure that resembles the one described in \cite{ham95}. Energy is constantly being redistributed to the mean flow into streaks and rolls. 

We note that \cite{ham95} analyzed $M(\alpha,0)$, which is invariant in the axial direction, and fundamental in $\theta$. \cite{ham95} states that this mode is responsible for rolls regeneration (\cite{ham95}, $\S$ 6), but in our simulations it evolves at faster time-scales and we do not see any positive or negative correlation between this mode and the two shown in Figure \ref{fi:tevo_streak_roll}(b). This indicates that we are not exactly seeing the SSP as studied by \cite{ham95}, but instead observe a cycle which has properties reminiscent of it. As a final remark, both the plots of figure \ref{fi:tevo_streak_roll} are realized at $Re=3.4 \cdot 10^4$, corresponding, as we already said, to the so called ``ultimate regime'', but we have observed almost identical behavior of the energies and modal velocities at smaller Reynolds number ($Re_s=5\times 10^3$, during the transition to this regime).

Additional evidence for a general similarity between the rolls and the SSP is given by a previous study performed by \cite{ost17} on decaying Taylor-Couette turbulence. In channel flow, the self-sustaining process regenerates itself with slow time-scales ($100 tU/d$). Once forcing is removed, the vortices would regenerate themselves slower and slower but with the same intensity, until the turbulence died out \citep{jim91,ham95}. In the decaying Taylor-Couette simulations of \cite{ost17}, similar phenomena could be observed. The cylinders were made stress-free at $tU/d=0$, thus removing the forcing from the rolls. The first life-stage of decay, attributed in the manuscript to the decay of the rolls lasted for a time-scale of $tU/d\approx 15-40$. The energy of the azimuthal (streamwise) velocity first decayed up to $tU/d\approx 10$, while the energy of the radial (wall-normal) and axial (span-wise) components did not start decaying up to $tU/d \approx 10$ and had a monotonic decay. This matches the findings of \cite{jim91,ham95}: the streamwise rolls are regenerated in full force in the first life-stage of decay until they completely die down and the flow enters the second life-stage of decay.

\subsection{The pinning of Taylor rolls}

\begin{figure}
\includegraphics[trim=5.7cm 0.7cm 5cm 0 0.7cm, clip, height=0.34\textwidth]{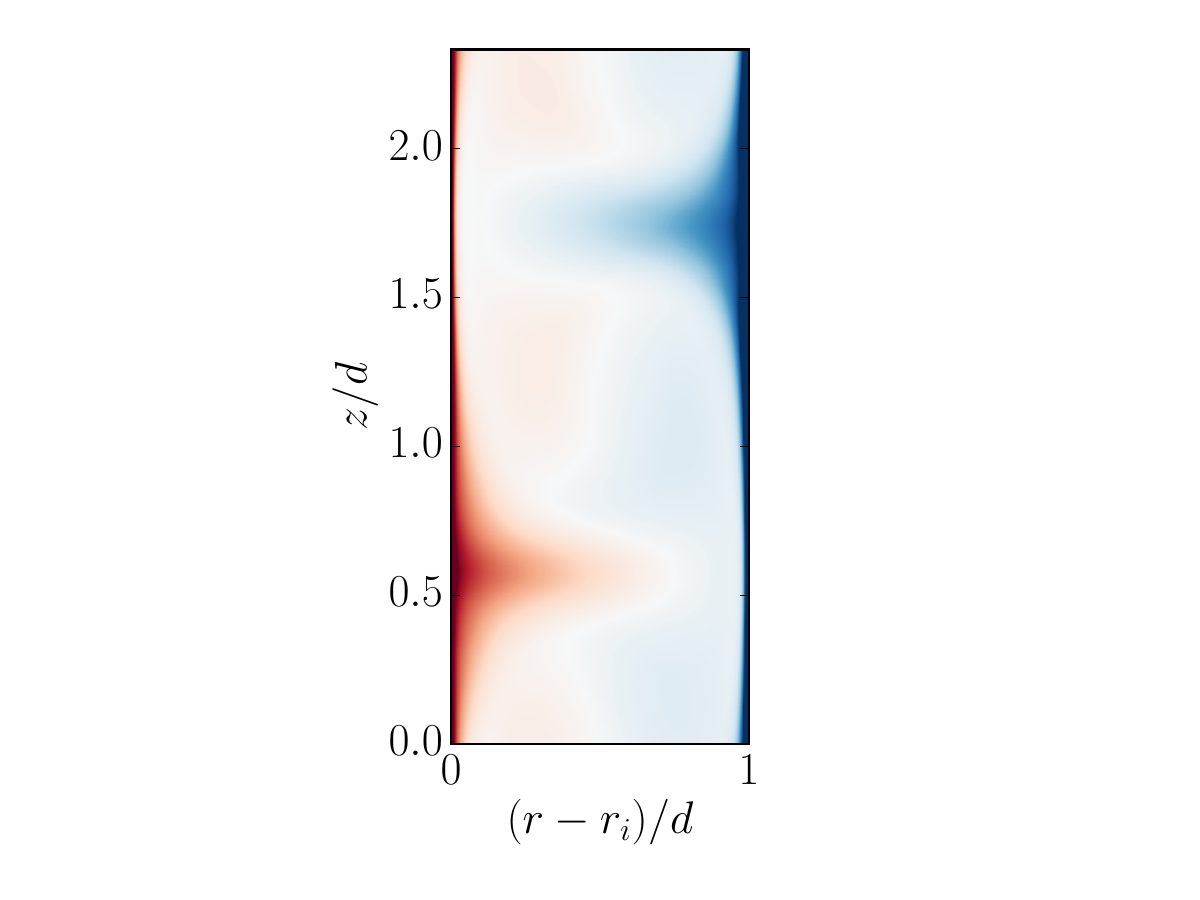}
\includegraphics[trim=8.6cm 0.6cm 0cm 0.4cm, clip, height=0.34\textwidth]{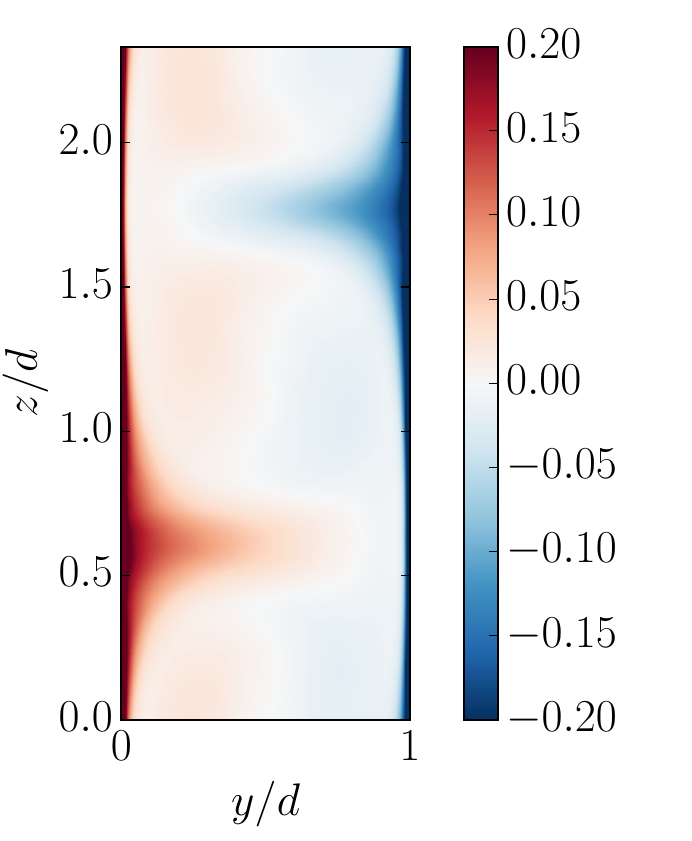}
\includegraphics[height=0.35\textwidth]{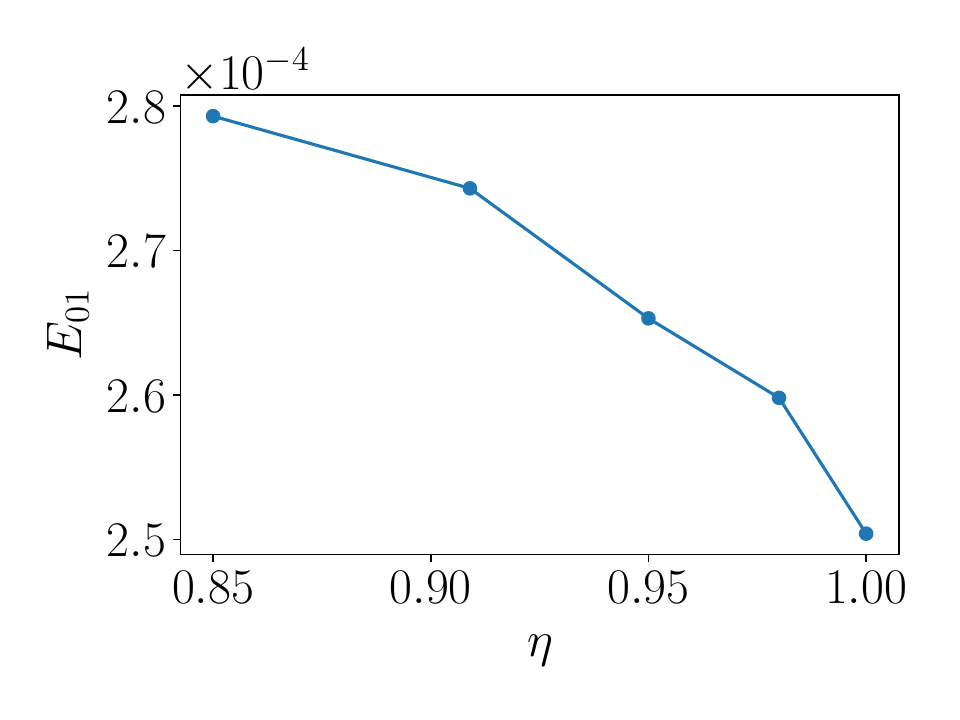}
\centering
\caption{Left: Mean azimuthal velocity for Taylor-Couette at $\eta=0.85$ and $R_\Omega=0.1$, $Re_s=3.4\cdot10^4$. Center: Mean streamwise velocity for plane Couette flow ($\eta \to 1$) at $R_\Omega=0.1$, $Re_s=3.4\cdot10^4$. Spanwise-pinned large-scale structures analogous to the turbulent Taylor rolls can be seen. Right: strength of the Fourier component associated to the turbulent Taylor roll at the mid-gap as a function of curvature for $R_\Omega=0.1$, $Re_s=3.4\cdot10^4$.} 
\label{fi:etatoone} 
\end{figure}

If the Taylor roll is reminiscent of the self-sustained process, which is axially pinned, what is causing the pinning? Is it the instabilities caused by curvature, or can rotation alone cause the pinning of the rolls? To answer this question we have simulated several narrow-gap radius ratios up to the limit of $\eta \to 1$, i.e.~the Plane Couette limit. By varying the curvature and keeping the other control parameters constant we can control for its effect. We set $Re_s=3.4\cdot 10^4$, in the ultimate regime \citep{ost14}. The non-dimensional rotation parameter $R_\Omega$ is set to $R_\Omega=0.1$, so we are considering weak anti-cyclonic rotation very close to the value for pure inner cylinder rotation at $\eta=0.909$, which is $R_\Omega=0.0909$. Analyzing the instantaneous and averaged azimuthal or streamwise velocity fields we find that large-scale structure analogous to Taylor rolls are present for all cases. The average and instantaneous flow-fields look very similar, so we expect them to be sustained by the same pattern of streaks that we have described previously. These facts are illustrated in the left and center panels of Figure \ref{fi:etatoone}. 

Moreover we have checked the magnitude of the rolls as curvature vanishes by looking at the  magnitude of the Fourier component associated to them in the $u_r$-$u_\theta$ mean at the mid-gap: the axisymmetric mode in the streamwise direction ($k_x=0$) and the axially/spanwise mode associated to the roll wavelength ($k_z = 2 \pi/\lambda_z$). This is shown in the right panel of Figure \ref{fi:etatoone}, where we can clearly see that the variations of this quantity are around $6\%$ between $\eta=0.909$ and $\eta=1$. This matches the intuition coming from the Navier-Stokes equations that the strength of forces due to curvature is measured by the $R_c=d/\sqrt{r_ir_o}$ parameter \citep{bra15}, which corresponds to $R_c\approx 0.095$ for $\eta=0.909$. This is further evidence that the nature of the rolls are similar in all the systems, and that their pinning is caused by rotation, and not curvature. Indeed, the Coriolis forces arising from solid body rotation, reflected non-dimensionally as $R_\Omega$, seem crucial in controlling the regions of parameter space in which rolls form.  

We note that while we use the word ``pinning'', this is a simplification. Coriolis forces do not break the vertical translation symmetry of the problem. Instead, their presence modifies the rolls and adds a feature that makes them more resistant to spanwise translation when compared to the $R_\Omega=0$ case. Large-scale structures are known to exist in non-rotating Plane Couette flow from the work of \cite{tsu06}. These were isolated in the simulations of \cite{avs14}, and found to consist of counter-rotating pairs of rolls with high vorticity in their boundaries. This pattern is similar to the  structure of the Taylor rolls seen in Figure \ref{fi:avgvelsvorts}. It is not curvature, but the addition of a mild anti-cyclonic rotation $R_\Omega=0.1$ which breaks the spanwise wandering of these structures, resulting in the fixed rolls. Proof of this is seen not only in our simulations, where the rolls are barely modified as the curvature vanishes, and correspond to the structures of \cite{avs14}, but also in the large-box rotating Plane Couette simulations of \cite{tob17} which show identical signatures for RPCF.

While in principle any rotating reference frame could be chosen to express a Coriolis force, obtaining a different expression for $R_\Omega$, the choice of this particular one in which the velocities at the cylinders are equal and opposite is the correct way to reproduce the transition between TC flow and RPCF \citep{bra15}. As the curvature increases, TC flow driven by pure inner cylinder rotation can be seen as TC flow driven by counter-rotating cylinders with an increasing solid-body rotation. The corresponding Coriolis force increases with increasing $\eta$, i.e. $R_\Omega(Re_o)=1-\eta$. With this Coriolis parameter $R_\Omega$, we can revisit older data to show that mild anticyclonic rotation, and not centrifugal instabilities, play a defining role for moderate values of curvature. By recalculating the parameters in terms of $R_\Omega$, we see that the $\eta=0.714$ ($R_c\approx 0.33$) experiments of \cite{hui14} show the signature of rolls between $R_\Omega \approx 0.1$ and $R_\Omega \approx 0.2$ when the cylinder counter-rotation is expressed as a Coriolis force, corresponding to the values discussed above. In the $\eta=0.5$ ($R_c\approx 0.71$) experiments of \cite{van16}, a fixed roll structure for counter-rotating cylinders was seen at $\omega_o/\omega_i=-0.2$, which corresponds to $R_\Omega=0.25$, but not for pure inner cylinder rotation corresponding to $R_\Omega=0.5$. This also serves to rationalize why the rolls vanish for pure inner cylinder rotation at $\eta=0.5$ ($R_\Omega=0.5$) and $0.714$ ($R_\Omega=0.286$), but not for $0.909$ ($R_\Omega=0.091$) as seen in \cite{ost14}. Mild anti-cyclonic Coriolis forces and not centrifugal forces are the primary cause of the pinning of structures in the ultimate regime.

\subsection{The effect of the domain size}

From the previous sections, it seems apparent that mild anticyclonic rotation causes some self-organization in shear-flows by pinning the self-sustained process in the span-wise (axial) direction. However, while the pinning seemed natural for experiments, due to the presence of end-plates, it is still unclear what the effect of box-size is on simulations. As mentioned in the introduction, it has been found by numerical simulation of \cite{lee18} that for plane Couette flow, a large computational box of $20 \pi \delta \times 2\delta \times 5\pi\delta$, where $\delta$ is half the distance between the walls, cause the pinning of flow structures. We note that this pinning was not observed either in the DNS simulations in a domain of size $20\pi\delta \times 2\delta \times6\pi\delta$ by \cite{avs14}, nor the $18\pi\delta\times2\delta\times8\pi$ domain of \cite{pir14}.

To probe the pinning of rolls, we decompose the velocity in a similar manner as in section 3.2 into mean flow and fluctuations. We return to the definition of streamwise averaged velocities $U(t,r,z)$ from Equation \ref{eq:axisym}, and we take a further step, as in \cite{MoserMoin87} by analyzing two derived quantities. First, we look at the axisymmetric azimuthal vorticity, which provides a measure of the circulation of the rolls:




\begin{equation}
\Omega_{\theta} = \frac{\partial U_r}{\partial z} -\frac{\partial U_z}{\partial r}.
\end{equation}
\noindent and second, we look at the axisymmetric axial anomaly, computed taking out the $\theta$ and $z$ average of the field from $U$, that was used by \cite{des18} (Eq. \ref{eq:straxi}) to identify streaks: 

\begin{equation}
U_s(t,r,z) = U(t,r,z) - \left\langle U(t,r,z) \right\rangle_{z},
\end{equation}
\\

\begin{figure}
\includegraphics[trim=4cm 1cm 6cm 0, clip, height=0.36\textwidth]{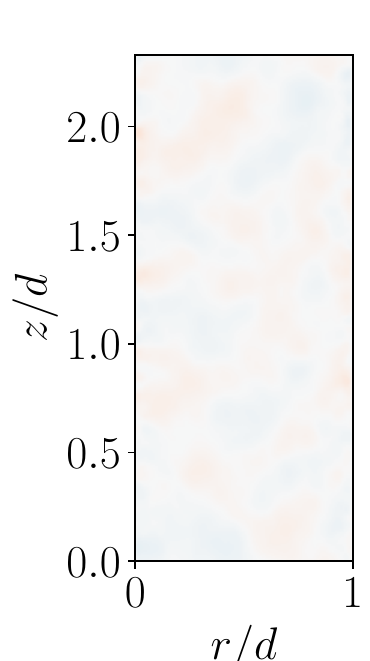}
\includegraphics[trim=6cm 1cm 1cm 0, clip, height=0.36\textwidth]{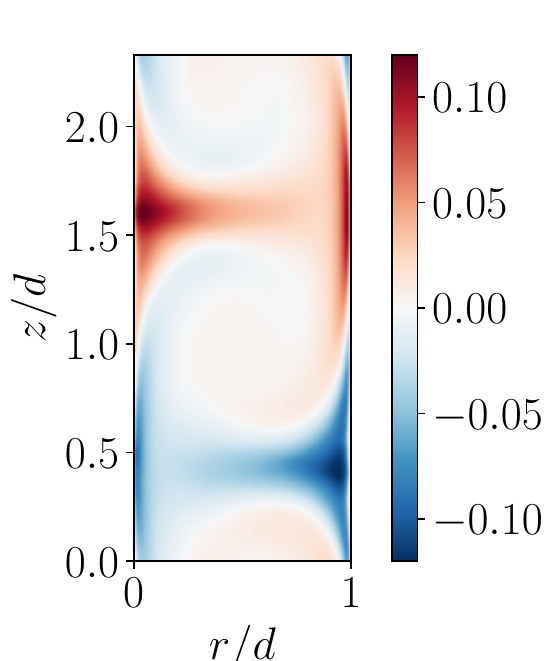}
\includegraphics[height=0.33\textwidth]{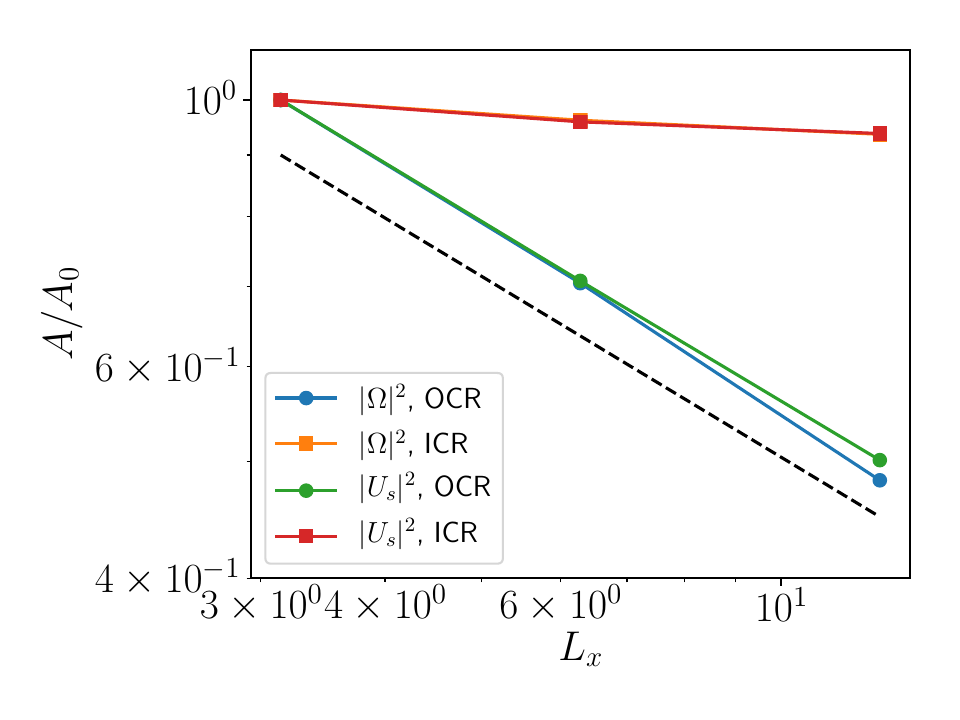}
\centering
\caption{Left and center: Mean azimuthal velocity anomaly ($U_s$) for outer cylinder rotation (left) and inner cylinder rotation (center) at $Re_s=3.4 \cdot 10^4$, $\eta=0.909$. Right: Normalized energy of the streaks and rolls as a function of the streamwise domain length at the mid-gap $L_x/d=2\pi (r_i+r_o)/(2dN)$ for the same values of $Re_s$ and $\eta$ for inner cylinder rotation (ICR) and outer cylinder rotation (OCR). The dashed line indicates the expected $A\sim L_x^{-1}$ behaviour for a structure-less flow where the energy comes from aliasing.}
\label{fi:boxsize} 
\end{figure}

We first compare two TC systems, with $R_\Omega=0.0909$, corresponding to pure inner cylinder rotation, with strong turbulent rolls, the other with $R_\Omega=-0.1$, corresponding to pure outer cylinder rotation, with fixed $Re_s=3.4\cdot10^4$ and $\eta=0.909$. As we previously discussed when analyzing $\Omega_{\theta}$ field, large scale axially pinned rolls are present only with a mild anticyclonic rotation, and so are a characteristic of the first system only, while the second does not present this feature at all. This is also reflected on the streak fields $U_s$, as we have found also there large-scale axially pinned structure, only for inner cylinder rotation, while they are completely missing in the other case. This is displayed in the left and central panels of figure \ref{fi:boxsize}. 

Moreover if we look closer at the central panel, we can see that the axial signature of the rolls has a well defined shape and an influence also in the bulk, analogous to what was seen for the azimuthal velocity in subsection \ref{subsec:LSE}. This is further proof that different mechanisms are taking place, and that Taylor rolls are part of an axially pinned process reminiscent of the SSP: the Coriolis forces are also pinning the streak fields, which in turn reinforces the rolls, keeping them fixed.

We now compare the effect of the azimuthal domain size on the two cases, to further prove that the rolls are not simply a product of aliasing, i.e. of large wavelengths mapped onto the streamwise invariant modes, caused by small domain extents. Even if recent studies have found that the computational domain (or box) size does affect the outline of the rolls \citep{ost15,ost16}, since the axial extent of the box plays a crucial role on the correlations and spectra of the velocity field, no size of box was able to unpin the rolls. Moreover larger boxes in the azimuthal extent allow for azimuthal wave-like patterns in the Taylor rolls to develop, which affects the statistics in the bulk region, but does not unpin them anyway. The question remains of how large is ``large enough''? Here, we attempt to rigorously quantify the effect of box-size to understand whether the presence of Taylor rolls is a product of numerics, since we are also imposing artificial rotational symmetry in the azimuthal direction. We can do this by systematically varying the azimuthal (streamwise) extent of the computational box. If the rolls are a product of aliasing, then, the energy of the rolls and streaks should decrease with box-size as $\sim 1/N$, as they are essentially a mean ($k_x = 0$). On the other hand, if they are physical, the energy should plateau to a constant with box-size.

We again use the comparison between pure inner rotating cylinder system and pure outer rotating one, as the last one, as we pointed out before, has a lack of roll structures. The simulations are performed with the same parameters, and varying the imposed rotational symmetry of the system between the values $N=5,10,20$. In the right panel of figure \ref{fi:boxsize} we show the streak and roll energies for both systems: for inner cylinder rotation we can see that even if the azimuthal extent of the domain change, the amplitude of rolls and streaks is almost constant; on the other hand for pure outer cylinder rotation the amplitude goes down with box-size as predicted for a mean mode.

As the last point we still want to understand if Taylor rolls are effectively fixed in the axial direction for all times or if they move about their position as a random process. For this reason, we decompose the flow in low-wavenumber roll modes and high-wavenumber fluctuations, in the spirit of the generalized quasi-linear approximation \citep{tob17}. We have already mentioned that the energy of the rolls is mainly contained in the axisymmetric ($k_x=0$) mode, and the axial mode associated to the base wavelength of the roll ($k_z = 2 \pi/\lambda_z$). By looking at the phase of the Fourier transform of the streamwise velocity $u_\theta$ at the mid-gap we can probe the extent of pinning. For the roll to move the phase of this mode, which we will denote from here as $\alpha$, has to shift with time.

\begin{figure}
\includegraphics[width=0.48\textwidth]{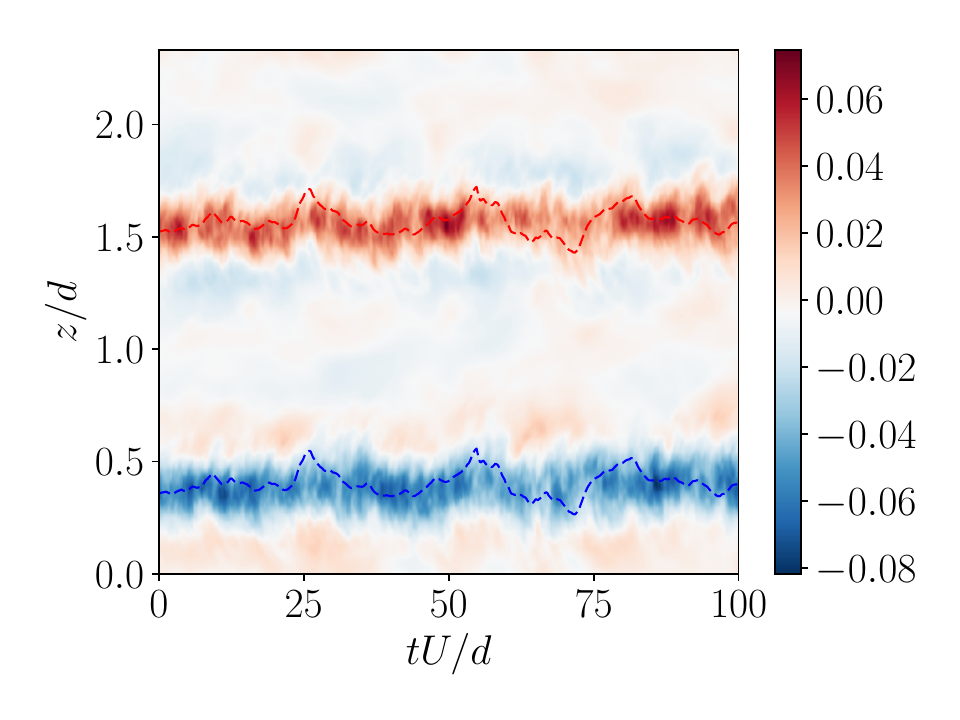}
\includegraphics[width=0.48\textwidth]{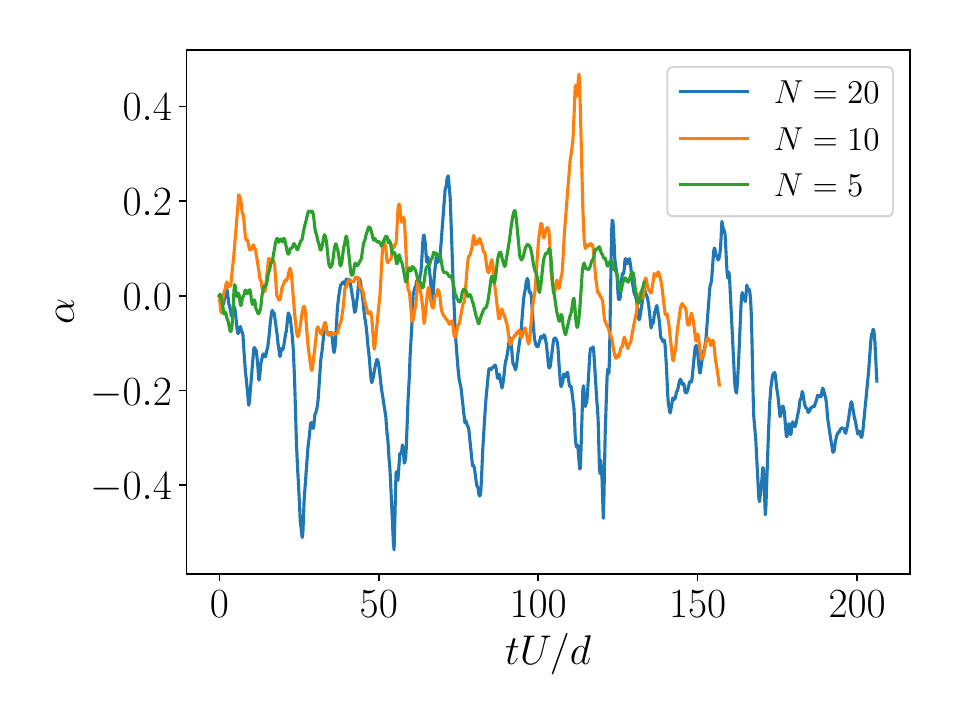}
\includegraphics[width=0.48\textwidth]{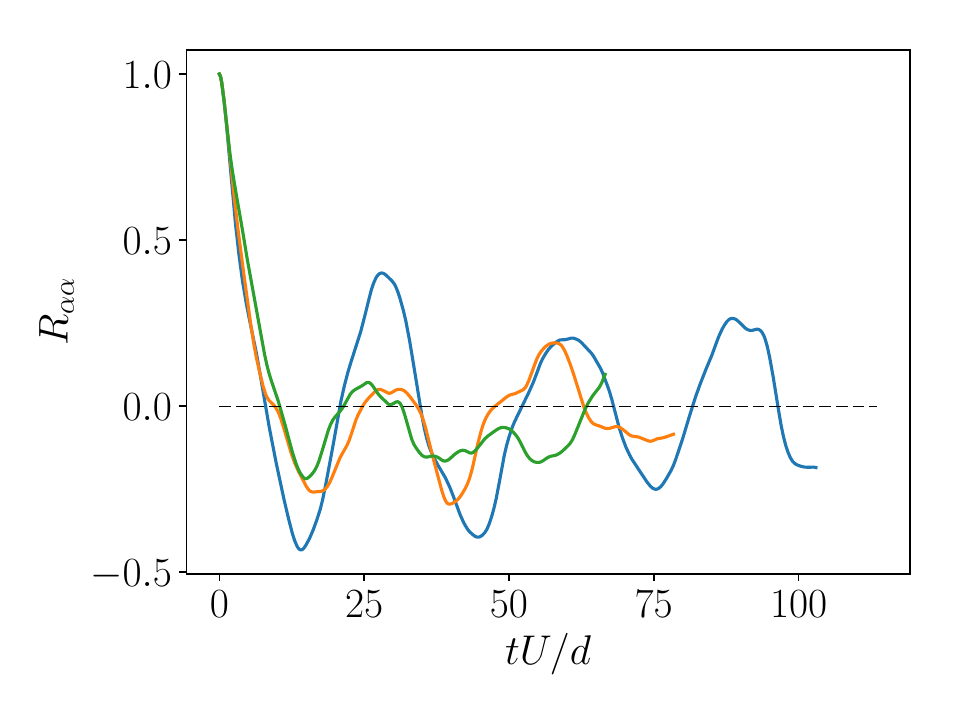}
\includegraphics[width=0.48\textwidth]{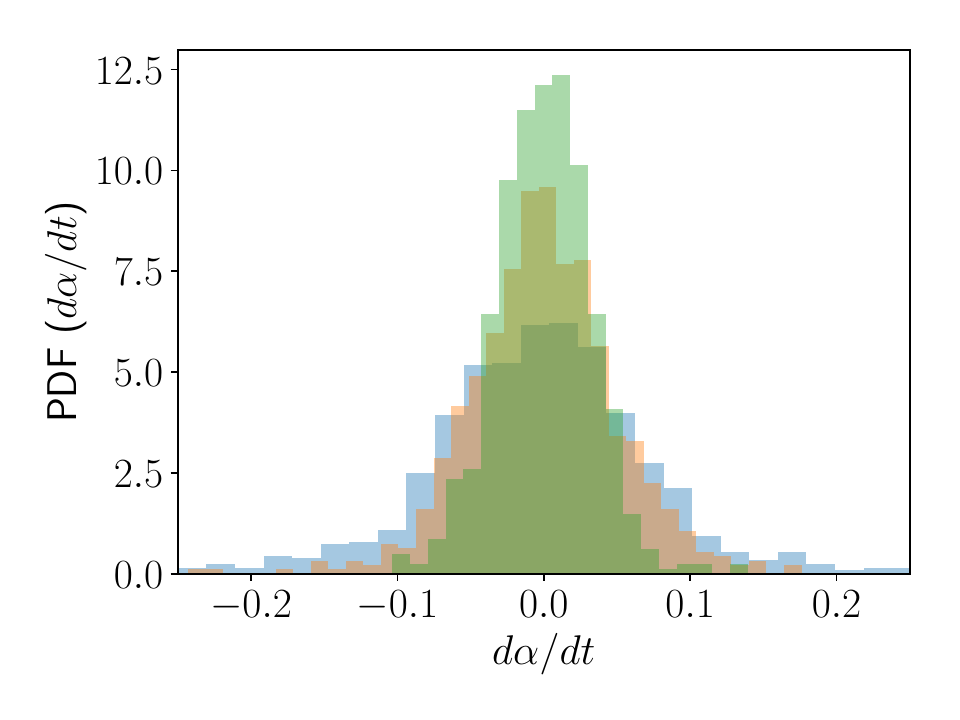}
\includegraphics[width=0.48\textwidth]{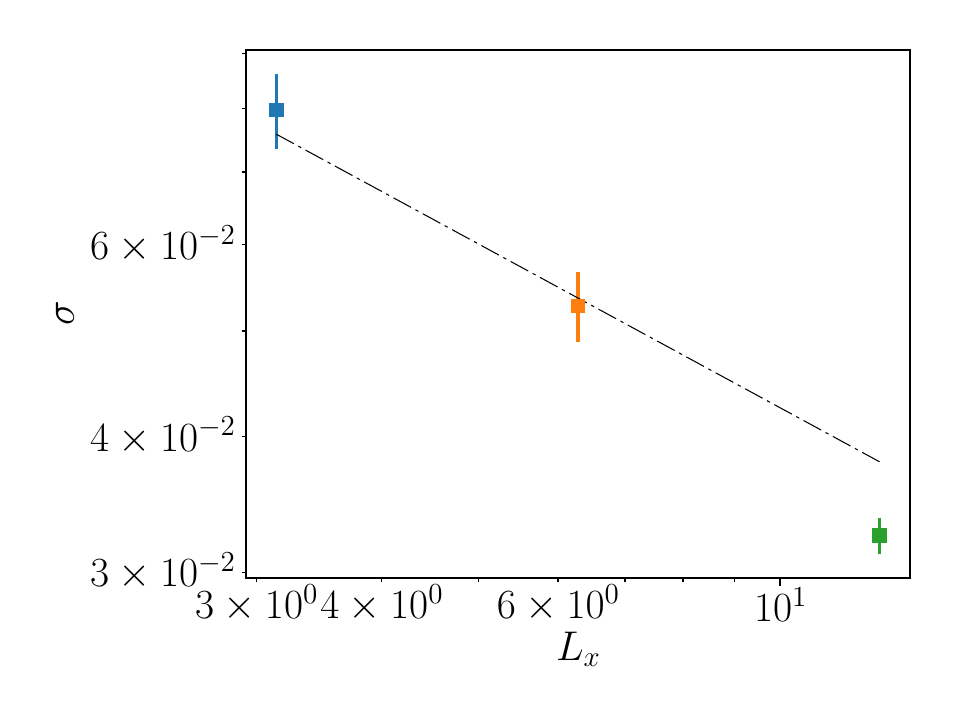}
\includegraphics[width=0.48\textwidth]{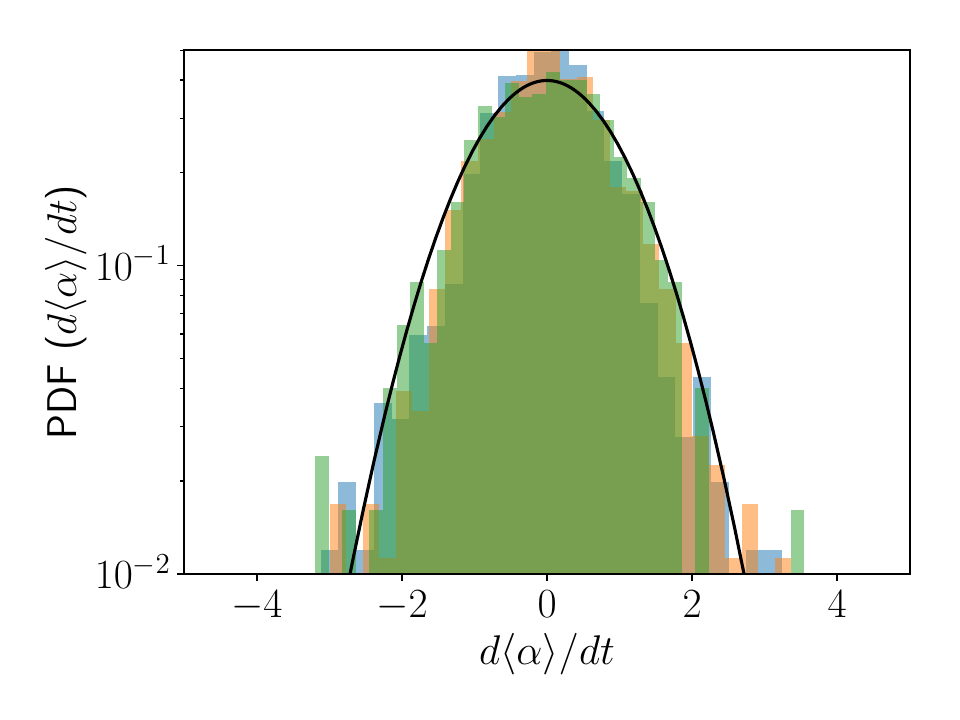}
\centering
\caption{Top left: Space-time (Hovm\"oller) diagram showing the streamwise-averaged azimuthal velocity at mid-gap for $N=20$ at $Re_s=3.4\cdot10^4$, $\Gamma=2.33$ and $\eta=0.909$. The position of the maximum and minimum velocities of the Taylor-roll Fourier mode is shown with a dashed line. Top right: Phase ($\alpha$) of the Fourier mode associated to the Taylor roll for three different azimuthal (streamwise) domain lengths . Middle left: Temporal auto-correlation of the phases for the three domains. Middle right: Probability distribution function for the phase velocity ($d\alpha/dt$) for the three simulations. Bottom left: Standard deviation of the phase velocity as a function of streamwise domain extent $L_x=20\pi/N$. The dashed line shows the proposed $1/\sqrt{L_x}$ behaviour. Bottom right: The three probability distributions of (d) normalized by their standard deviation with a Gaussian distribution overlaid.}
\label{fi:alpha} 
\end{figure}

In the top left panel of figure \ref{fi:alpha} we show a visualization of the roll movement, with the predicted peak from the phase of the Fourier component superimposed. The method tracks relatively well the position of the minimum and maximum velocity. We note that an implicit assumption we make when using the Fourier mode, is that each of the two rolls behaves in a symmetric manner, which might not be strictly the case. 
 
On the top right panel we show the time evolution of the phase for pure inner cylinder rotation cases at $Re_s=3.4\cdot10^4$ and $\eta=0.909$, with a varying imposed rotational symmetry between the values of $N=5,10,20$. The +movement of the rolls can be seen to become smaller for increasing box-size. We also show the temporal autocorrelation of the roll phase in the middle left panel of \ref{fi:alpha}, where it can be seen that the spatial changes decorrelate in time, with the decorrelation time increasing as the streamwise extent of the domain increases. Thus the drifts become both longer in time and shorter in space with increasing domain size.

Furthermore, we can measure the amplitude of the phase change (i.e.~the speed of the drift of the roll) $d\alpha/dt$, and calculate the probability distribution function (p.d.f.) of this quantity. This is shown in the middle right panel. The drifts are approximately symmetric, as could be expected by the symmetries in the system. In the bottom right panel, we show the three curves normalized by their standard deviation and compared to a standard Gaussian distribution. The three curves can be seen to approximate the Gaussian distribution with their standard deviation, i.e. the size of the spatial drifts, decreasing with increasing domain size.

It thus seems that the spatial drifts not only become shorter, but they also become more correlated with increasing domain size. This can be reasoned as follows: all non-linear interactions are triadic interactions of low or high wavenumber modes. A force that could shift the phase has to come from the interaction of two high-wavenumber fluctuations, that is, from a triadic interaction of the form $\widehat{u}'(k_1),\widehat{u}'(k_2) \to \widehat{U}(k_1 + k_2)$. If we now assume that fluctuations are random, and centered at zero, using the central limit theorem we expect that the mean of all triadic interactions affecting the roll behaves like a Gaussian. We also expect that an increase in the computational box, which adds more modes to the averaging, makes the expected value of the fluctuations approach zero. In the bottom left panel of Figure \ref{fi:alpha} we show the standard deviation of the drift, which approaches zero even faster than $1/\sqrt{L_x}$, that is what would be expected from a simple random process. This is an indication that the phenomenon is more complicated than what could be predicted from this simple analysis, but anyway the pinning process holds and works stronger than our expectation.

Overall, the results verify reasonably well our expectations. We have shown that rolls are moving, but this movement depends on the box-size: the larger the domain the smaller the spatial variations. Computational effects cannot be completely avoided. Indeed, boxes which are very large might pin structures by the averaging out fluctuations over long spatial extents. 

\section{Summary and conclusions}\label{sec:conclusion}

Several DNS simulations have been performed in a large parameter range of Taylor-Couette system in order to better understand the high-Reynolds number behaviour of the large-scale structures known as Taylor rolls. We have shown that, at a fixed radius ratio $\eta=0.909$, these structures become more and more empty of vorticity in the bulk as the Reynolds number increases, while the behaviour of the streamwise velocity remains somewhat similar, leaving behind the characteristic imprint in the axial direction on both instantaneous and mean fields. In a way similar to the rolls, large-scale streaky structures arise in the azimuthal (streamwise) direction. At high Reynolds numbers the roll/streak pairs can be understood in the context of a spatially localized instance of a process reminiscent of the self-sustained process of shear flows: the primary mechanism for the generation is not the centrifugal instabilities caused by rotation and curvature, as previously thought \citep{ost16}. Instead, streaks are generated by the redistribution of shear stress by the rolls, and their instability sustains the evolution of the same rolls. For a narrow gap system, the  self-sustained process is pinned under the presence of moderate anti-cyclonic rotation. The pinning is not due to destabilizing centrifugal effects, as the process remains pinned in the plane Couette limit value $\eta \to 1$, i.e.~for no curvature. The $R_\Omega$ parameter range where the rolls are pinned matches previous experimental results from large-gaps between $\eta=0.5-0.714$. Finally, we have shown that the rolls are not a product of aliasing due to small computational domains, and their position appears to change, governed by a random process with long time-scales. The smaller the domain, the larger the change in position of the rolls, because the average of the triadic interactions between fluctuations deviates more from zero. Further investigation should look at the role of the rotation parameter $R_{\Omega}$ on other important parameters, as torque, and on the velocity field, in order to understand when Taylor rolls have an optimal flow, and the effects that rotation and curvature have on large scales, whether alone or coupled.

\emph{Acknowledgments:} We thank B. Eckhardt, B. Farell and P. Ioannu for fruitful discussions. We thank the Center for Advanced Computing and Data Science (CACDS) at the University of Houston for providing computing resources and we also acknowledge PRACE for awarding us access to MareNostrum IV, based in Spain at the Barcelona Supercomputing Center (BSC) under PRACE project number 2017174146.

\bibliographystyle{jfm}
\bibliography{tc_literatur}

\end{document}